\documentclass[prb,twocolumn,amsmath,amssymb,aps,superscriptaddress,eqsecnum,floatfix]{revtex4-1}
\usepackage{graphicx}
\usepackage{dcolumn}
\usepackage{bm}
\usepackage[breaklinks]{hyperref}
\usepackage{subfigure}

\def\2DEG{two-dimensional electron gas~}
\def\1DEG{one-dimensional electron gas~}

\def\YBCO{YBa$_2$Cu$_3$O$_{6+x}$}

\def\C60{A$_x$C$_{60}$}

\def\sign{\textrm{sign}}

\def\SROtwo{ Sr$_{3}$Ru$_{2}$O$_{7}$}

\def\BSCCO{Bi$_2$Sr$_2$CaCu$_2$O$_{8+\delta}$}

\def\HgCu3{HgCa$_2$Cu$_3$O$_{8+y}$}
\def\HgCu4{HgBa$_2$Ca$_3$Cu$_4$O$_{10+y}$}
\def\TlCu{Tl$_2$Ba$_2$CuO$_{6+\delta}$}
\def\TlCu3{Tl$_2$Ba$_2$Ca$_2$Cu$_3$O$_{10+y}$}
\def\TlCu4{Tl$_2$Ba$_2$Ca$_3$Cu$_4$O$_{12+y}$}

\def\BiCu3{Bi$_2$Sr$_2$Ca$_{2}$Cu$_3$O$_y$}

\def\8LSCO{La$_{1.88}$Sr$_{.12}$CuO$_4$}
\def\110LNSCO{La$_{1.5}$Nd$_{0.4}$Sr$_{0.1}$CuO$_{4}$}
\def\stage4LCO{La$_{2}$CuO$_{4+\delta}$}
\def\Y248{YBa$_2$Cu$_4$O$_8$}

\def\NbSe2{NbSe$_2$}
\def\TaSe2{TaSe$_2$}
\def\TiSe2{TiSe$_2$}
\def\NaCoOH2O{Na$_{0.3}$CoO$_{2y}$H$_2$O}
\def\MgB2{MgB${}_2$}

\def\FeAs122{Ca(Fe$_{1-x}$Co$_x$)$_2$As$_2$}

\begin{document}

\title{Nematic quantum phase transition of composite Fermi liquids in half-filled Landau levels and their geometric response}

\author{Yizhi You}
\affiliation{Department of Physics and Institute for Condensed Matter Theory, University of Illinois at Urbana-Champaign, 
Illinois 61801-3080, USA}
\affiliation{Kavli Institute for Theoretical Physics, University of California Santa Barbara, Santa Barbara, California 93106}
\author{Gil Young Cho}
\affiliation{Department of Physics, Korea Advanced Institute of Science and Technology, Daejeon 305-701, Korea}
\affiliation{Department of Physics and Institute for Condensed Matter Theory, University of Illinois at Urbana-Champaign, 
Illinois 61801-3080, USA}
\author{Eduardo Fradkin}
\affiliation{Department of Physics and Institute for Condensed Matter Theory, University of Illinois at Urbana-Champaign, 
Illinois 61801-3080, USA}

\date{\today}
\begin{abstract}
We present a theory of the isotropic-nematic quantum phase transition in the composite Fermi liquid arising in  half-filled Landau levels. We show that the quantum phase transition between the isotropic  and the nematic phase  is triggered by an attractive quadrupolar interaction between electrons, as in the case of conventional  Fermi liquids. We derive the theory of the nematic state and of the phase transition. This theory is based on the  flux attachment procedure which maps an electron liquid in half-filled Landau levels into the composite Fermi liquid close to a nematic transition.    
We show that the local fluctuations of the  nematic order parameters act as an effective dynamical metric interplaying with the underlying  Chern-Simons gauge fields associated with the flux attachment. Both the fluctuations of the Chern-Simons gauge field and the nematic order parameter can destroy the composite fermion quasiparticles  and drive the  system  into a non-Fermi liquid state. The effective field theory for the isotropic-nematic phase transition is shown to have $z = 3$ dynamical exponent due to the Landau damping of the dense  Fermi system. We show that there is a Berry phase type term which governs the effective dynamics of  the nematic order parameter fluctuations, which can  be interpreted as a non-universal  ``Hall viscosity" of the dynamical metric. We also show that the effective field theory of this compressible fluid  has a Wen-Zee-type term. Both   terms  originate from the time-reversal breaking fluctuation of the Chern-Simons gauge fields.  We present a perturbative (one-loop) computation of the Hall viscosity and also show that this term is also obtained by a Ward identity.  We show that the topological excitation of the nematic fluid, the disclination,  carries an electric charge. We show that a resonance observed in  radio-frequency conductivity experiments can be interpreted as a Goldstone nematic mode gapped by  lattice effects.
\end{abstract}

\maketitle

\section{Introduction and Motivation}
\label{sec:intro}

Electronic nematic phases have been a focus of attention during the past few years in several areas of quantum condensed matter physics.\cite{Fradkin-2010} An electronic nematic  is a state of a strongly correlated electronic system in which rotational invariance is broken spontaneously without breaking translation symmetry. Unlike their classical liquid crystal cousins,\cite{chaikin-1995,degennes-1993} whose tendency to exhibit orientational order can be traced back to the microscopic cigar-shaped nature of the constituent nematogen molecules, an electronic nematic phase arises from the self-organization of electrons in a strongly correlated material.

The electronic nematic state belongs to a class of phases of strongly-interacting quantum-mechanical electronic matter, known as electronic liquid crystal states,\cite{kivelson-1998,Fradkin-1999} which are characterized by the spontaneous breaking of the spatial symmetries of a physical system. Electronic nematic phases have by now been discovered experimentally in many different systems ranging, among others, from high temperature cuprate superconductors, such as in {\YBCO}\cite{ando-2002,hinkov-2007,Daou-2010} for a broad range of doping levels, and in underdoped {\BSCCO},\cite{Lawler-2010} to iron-based superconductors such as {\FeAs122},\cite{Chu-2012} and also to the bilayer ruthenate {\SROtwo}.\cite{borzi-2007}

However, the first and to this date the most spectacular experimental evidence for an electronic nematic state was  discovered in two-dimensional electron gases (2DEG) in high magnetic fields in the middle of the second, $N=2$, Landau level (and higher), in regimes in which the 2DEG is compressible and the fractional quantum Hall (FQH) effect is not observed.\cite{lilly-1999,du-1999} In these experiments, longitudinal and Hall transport measurements were made in the center of the Landau level for Landau levels $N \geq 2$. It was  found that the longitudinal transport properties exhibit a strong spatial anisotropy (with a ratio of resistances as large as 3,500 in the cleanest samples at the lowest temperatures, originally down to $T \gtrsim 25$ mK). This anisotropy has a fairly rapid increase at a temperature $T \simeq 65$ mK from a nominal anisotropy of a fraction of a percent at $T \sim 1$ K. Importantly, in this regime the I-V curves are linear at low bias and, hence, do not show any signs of translation symmetry breaking, e.g. no  threshold electric fields, characteristic for a charge-density-wave ground state, were ever detected in this regime. 
 In contrast, in the same samples and at the same temperatures, a reentrant integer quantum Hall plateau is observed away from the center of the Landau level, and, in this regime, an extremely sharp threshold electric field is  seen,  with a sharp onset of narrow-band noise for larger electric fields.\cite{cooper-2002,cooper-2003,Fradkin-2010} Nevertheless, these experiments were originally interpreted as evidence of a striped ground state, an interpretation still  used in the literature.
 
 Hence, in the compressible anisotropic regime, down to the lowest temperature accessible in the experiments (which currently go down to about $10$ mK), the 2DEG behaves as a compressible charged fluid with a large anisotropy which onsets below a well defined temperature. This behavior strongly suggested that there is a (thermal) phase transition of the 2DEG, rounded by a very weak native anisotropy (with a characteristic energy scale estimated to be $\sim 3-5$ mK, whose microscopic origin has remained unclear\cite{Pollanen-2015}) to a low-temperature electronic nematic state.\cite{Fradkin-1999} The nematic nature of the state was verified by detailed fits of the transport anisotropy data to classical Monte Carlo simulations of the thermal fluctuations of nematic order.\cite{Fradkin-2000,cooper-2001}

Subsequent experiments in 2DEGs in quantum wells, earlier  by tilting the magnetic field,\cite{lilly-1999b,pan-1999} and, more recently, by the application of hydrostatic pressure in the absence of an in-plane magnetic field,\cite{Samkharadze-2015} have revealed the existence of a complex phase diagram in which compressible nematic phases were found even in the first Landau level, $N=1$,  competing with the famous, presumably non-Abelian, paired, FQH state at filling fraction $\nu=5/2$. More recent tilted field experiments have also revealed the existence of an incompressible nematic FQH state in the $N=1$ Landau level at filling fraction $\nu=7/3$, competing with the isotropic Laughlin-like FQH state at that filling fraction.\cite{Xia2010,Xia2011} No nematic state has ever been reported in the lowest, $N=0$, Landau level.

Early Hartree-Fock theories of the 2DEGs near the center of the Landau level, for Landau level index $N$ large enough, have predicted a stripe-like ground state, i.e. a compressible state in which the electron density is spontaneously modulated along one direction.\cite{moessner-1996,koulakov-1996,koulakov-1996b} For this reason,  the anisotropic states in the compressible regimes in Landau levels $N\geq 2$ were originally referred to as striped states. Most   microscopic theories for the anisotropic state at $\nu=9/2$ (and in higher Landau levels) were built on this proposal.\cite{macdonald-2000,Stanescu-2000,barci-2002,Fogler2002,Cote-2000,Lopatnikova-2001,lawler-2004} The resulting picture of the stripe state is an array of ``sliding'' Luttinger liquids.\cite{emery-2000,vishwanath-2001,Sondhi-2001,Fertig-1999} 

On the other hand, an exact diagonalization study by Rezayi and Haldane\cite{rezayi-2000}  for a system of up to 16 electrons  for half-filled Landau levels in a toroidal geometry  gave strong evidence for both a paired FQH state and a stripe-like state as a function of the effective interactions in the Landau level. We should note that in such small system (and in a toroidal geometry) finite-size effects can blur the distinction between a stripe state and a nematic state, but it is an evidence for at least short-range stripe order.

Interest in nematic quantum Hall states attracted renewed attention  after the  experimental discovery of an incompressible nematic phase inside the fractional quantum Hall state in the $N=1$ Landau level at filling fraction $\nu=7/3$ by Xia and coworkers.\cite{Xia2010,Xia2011} This state has been studied theoretically by several groups.\cite{Mulligan2010,you2013,you2014, Haldane2009, Haldane2011,Yang2013, Maciejko2013}  These studies have revealed that  nematic fluctuations are intimately related to the geometric response of the quantum Hall fluid and, in particular, to the Hall viscosity. The incompressible nature of nematic fractional quantum Hall states strongly constrains the behavior of the nematic fluctuations and largely determines the structure of the effective behavior at low energies. These studies have also shown that the nematic transition inside the FQH state is triggered by a softening and condensation of the stable collective mode of the FQH fluid, the Girvin-MacDonald-Platzman (GMP) mode, at zero momentum. 

The close vicinity of nematic order of a compressible state or a FQH state (which is hence incompressible) strongly suggests that the nature of the effective interactions in the 2DEG in Landau levels $N\geq 1$ favor both paired and nematic ordered states. 
In this context, it is  surprising that, in spite of all the work on nematicity in the incompressible state, there has been almost no work on the  compressible nematic state for more than a  decade. Aside from  the notable semi-phenomenological theory of the quantum Hall nematic of Radzihovsky and Dorsey,\cite{Radzihovsky2002}  based both on a microscopic theory and on quantum hydrodynamics, and of the work of  Wexler and Dorsey,\cite{wexler-2001} who made estimates of the dislocation-unbinding transition of a quantum Hall stripe state to a quantum Hall nematic based on the Hartree-Fock theory of the stripe state, except for  a variational Monte Carlo wave function study by Doan and Manousakis,\cite{doan-2007} the compressible nematic state has not been studied. 

 In principle there are two logical pathways to reach a nematic phase by a quantum phase transition: a)  by quantum melting of a stripe phase,  or b) by a (Pomeranchuk) instability of  an isotropic Fermi liquid type state. Although the close vicinity of the isotropic compressible Fermi liquid state to the observed nematic state in Landau levels $N \geq 1$ suggests that the latter may be a suitable starting point, the fact that exact diagonalization studies find local stripe order\cite{rezayi-2000} suggests that the actual physics is likely to lie somewhere  in between these two regimes. Also, it is possible that the state is nematic above some critical temperature while the ground state may be a stripe phase. However, the fact that there is  strong evidence for rotation symmetry breaking but not of translation symmetry breaking, down to the lowest  experimentally accessible temperatures, suggests that the ground state may be a nematic state (perhaps close to a quantum phase transition to a stripe phase.)
 
 The purpose of this paper is to  develop a theory of the compressible nematic state of the 2DEG in large magnetic fields. Throughout this work we will use the mapping of electrons in Landau levels to composite fermions in the same Landau levels but now coupled to a Chern-Simons gauge field,\cite{Lopez-1991,Fradkin-1991} i.e. a flux attachment transformation. At a formal first-quantized level this mapping is an exact identity. However the resulting theory has no small parameter and in practice a mean field theory, the average field approximation, must be used. For FQH states, which have a finite energy gap already at the level of the mean field theory, this approach has been shown to yield exact  predictions of the universal properties  of the fractional quantum Hall fluid, including the Hall conductance, the charge and statistics of the excitations, degeneracy on a torus, and the Hall viscosity and related geometric responses.\cite{Fradkin-1991,Cho-2014,Gromov-2015}  On the other hand, here we will be interested in compressible phases  which do not have a gap (by definition) and hence the theory is not as well controlled as in the FQH regime. As our starting point  we will consider  the isotropic Fermi liquid state of Halperin, Lee and Read\cite{Halperin-1993,Read-1994} (HLR)  of composite fermions,\cite{Jain1989}  which is based on the same mapping, and look for a Pomeranchuk quantum phase transition to a nematic state.   This is a natural point of view which has been used extensively as a description of nematic Fermi fluids.\cite{Oganesyan-2001} However, the HLR Fermi liquid is a non-Fermi liquid to begin with which makes the  application of these ideas not straightforward.
 
 At the level of mean field theory,  the HLR (or Jain) Fermi liquid states are the limiting state of the FQH states in the  Jain sequences\cite{Jain1989} with filling fraction $\nu=p /(2np\pm1)$, where $p$ and $n$ are two non-negative integers. In Jain's picture, a FQH state of electrons can be viewed (in mean field theory) as an integer quantum Hall state with filling fraction $\pm p$ of composite fermions, which is made of gluing $2n$ fluxes to every electron, in  a partially screened magnetic field $B/(2np\pm 1)$. In the compressible limit, $|p| \to \infty$, the effective magnetic field felt by the composite fermions vanishes (on average) or, equivalently, the effective charge of the fermions vanishes in the same limit. In this regime, the charge-neutral composite fermions fill a Fermi sea and form a (composite) Fermi liquid.\cite{Jain-1992} This simple picture, and its subsequent extensions, has given a successful description of numerous experiments in the compressible regime.\cite{Willett-1990,Kang-1993,Goldman-1994,Jain-2007}  Also, a current picture of the microscopic origin of the non-abelian paired state in the first, $N=1$, Landau level, at filling fraction $\nu=5/2$, is a paired state in the $p_x+ip_y$ channel of composite fermions.\cite{Park-1998,Read-2000} It is then natural to look for a similar quantum transition to a compressible nematic state from the HLR state.

 In Ref. [\onlinecite{you2014}] we worked out a theory of the incompressible nematic state in a fractional quantum Hall state using the flux attachment via fermion Chern-Simons gauge theory\cite{Lopez-1991}  as a quantum phase transition in a Laughlin FQH state. Much as in the case of the nematic transition in a conventional Fermi liquid,\cite{Oganesyan-2001} we showed that the quantum phase transition (the Pomeranchuk instability) can be caused by an effective quadrupolar interaction of among the electrons if it becomes sufficiently attractive. Furthermore, we showed that the electrons feel fluctuations of the nematic order parameter  as an effective dynamical metric field. A direct consequence of this coupling is that the effective Lagrangian of the nematic fluctuations has a time-reversal breaking parity-odd Berry phase term, which is closely related to (but not equal to) the Hall viscosity. Due to this parity odd term, the quantum critical theory has dynamical exponent $z=2$. There are also Wen-Zee-like term in terms of the ``effective" dynamical metric, i.e., nematic order parameters. In the nematic phase, there is a topological soliton, the nematic disclination, which in this fluid carries an (unquantized) electrical charge. The resulting effective field theory of the nematic fractional quantum Hall state obtained by this approach has the same structure and properties as the one proposed on symmetry grounds by Maciejko and coworkers.\cite{Maciejko2013}
  
 The theory of the compressible nematic state that we will present here is naturally connected to our earlier work on the incompressible nematic FQH  state. 
 Thus, we will represent the problem of the half-filled Landau level and, in fact, for all  the compressible limiting states of the Jain sequences, at filling fraction $\nu=1/(2n)$, as a system of composite fermions minimally coupled (i.e. in gauge-invariant way) to both the external electromagnetic field $A_\mu$ and to the statistical gauge field $a_\mu$ which implements the  flux attachment. Hence, here too, the action also includes a Chern-Simons term (with a suitable coefficient). 
 
 Also, in addition to the  Coulomb interaction, which only involves a coupling of the local densities, as in the case of a Fermi liquid,\cite{Oganesyan-2001} we will also include an attractive interaction in the quadrupolar channel, i.e. an attractive interaction between the nematic densities. The coupling constant of this quadrupolar interaction is nothing but the  $F_2$ Landau parameter of a Fermi liquid. By  gauge invariance, the quadrupolar coupling of the fermions also involves both the gauge fields, $A_\mu$ and $a_\mu$, since the nematic densities are bilinear of the Fermi fields which necessarily involve spatial  derivatives. 
 
 We will not attempt here to provide a microscopic derivation of the value (and sign) of the effective quadrupolar interaction. Nevertheless, it is well known that the effective interactions  of composite fermions are quite different than those of electrons,\cite{Jain-2007} and depend on the Landau level index, as well as on properties of the heterostructure (or quantum well) which define the 2DEG. In addition, the estimate of values of Landau parameters, which is notoriously difficult even for conventional Fermi liquids, is much harder in the case of composite fermions. The  currently available numerical estimates\cite{Lee-2015}  for $F_2$  obtain values  that for $N \geq 1$ are very close to $-1$.
 
 However, the physics of the compressible state is actually quite different from the FQH states, and the extension of this theory to the compressible state involves several problems.  One is the lack of a small expansion parameter to control the theory. In the incompressible FQH states this is not a serious problem provided that one focused only on the long distance and low energy regime where it behaves as a topological fluid. This simplification is absent in the compressible state since it is gapless. Even though at the level of  mean field theory the state is predicted to be  a Fermi liquid, the coupling of the fluctuations of the statistical gauge field turns the (mean-field) HLR state into a non-Fermi liquid. Thus, already at the leading  perturbative order, imaginary part of the the composite fermion self-energy $\Sigma''(\omega)$  overwhelms the real part,\cite{Halperin-1993,Nayak-1994} i.e. $\Sigma''(\omega) \propto |\omega|^{2/3}$ for short-range interactions although it is milder for Coulomb interactions, $\Sigma'' (\omega) \propto |\omega|$ ( a ``marginal'' Fermi liquid), and the composite fermion quasiparticles become ill-defined. In addition, a calculation of the Landau parameters $F_l$ for the HLR theory\cite{Stern-1995} revealed that, for all angular momentum channels $l \geq 1$, all the Landau parameters are equal to the Pomeranchuk value, $F_l=-1$. Therefore, the quasiparticle picture breaks down and even the relic of a Fermi surface appears to be prone to instabilities (such as a nematic instability). Current numerical estimates  in the half-filled Landau level yield negative  values of $F_2$ and close to   the Pomeranchuk instability value.\cite{Lee-2015} Already the conventional theory of the nematic transition is non-trivial since at the Pomeranchuk point the Fermi liquid breaks down, and in the HLR state the Fermi liquid picture has been already broken down due to the coupling to the fluctuating gauge field. Nevertheless, properties of the fluid determined by gauge-invariant currents and densities are well behaved, in the sense that they are free of infrared singularities, although the  results are at best qualitative since the theory does not have a small parameter.\cite{kim1994gauge,Shankar-1997} 
 
 A further complication is the lack of particle-hole symmetry in the HLR theory.\cite{Kivelson-1997} This problem  is the focus of intense  current work.\cite{Son-2015,Barkeshli-2015,Wang-2015,Geraedts-2015,Metlitski-2015,Kachru2015,Mulligan2016,Mross2015,Murthy2016} It has also been a focus of attention in the theory of the paired (Pfaffian) FQH state.\cite{Lee-2007,Levin-2007} Particle-hole symmetry in the half-filled Landau level can exist only in the absence of Landau level mixing and if only quadratic interactions are allowed. On the other hand, although the flux attachment transformation is an exact mapping, at the level of  the average field approximation there is a large reorganization of the Hilbert space which involves a large mixing of Landau levels.  For this reason  the Jain wave functions are projected onto the Landau level. However, in the field theory approach there is no such projection. 
 
In the incompressible FQH states the effects of Landau level mixing become negligible at long distances and at low energies provided that the quantum fluctuations (``one-loop'' or ``RPA'') are included, as a consequence of incompressibility, Galilean and gauge invariance.\cite{Lopez-1992,Lopez-1993} The correct universal properties, encoded in the effective topological field theory, of the FQH states are reproduced  only  after  these leading quantum corrections are included. These quantum corrections at long distance and at low energies turn the composite fermions into anyons with fractional charge and fractional statistics and, in this sense, there are no composite fermions in the spectrum of states of the FQH fluids. On the other hand, because of the absence of a small parameter, the theory yields quantitatively incorrect values for dimensionful (and non-universal) quantities, often by significant amounts although improvements have been made.\cite{Shankar-1997} These problems become more complex in the case of the HLR state and, for this reason, theories of the compressible state projected onto the Landau level have been introduced.\cite{Pasquier-1998,Read-1998} These theories are technically more complex and non-local, and are only qualitatively understood.
  
 In what follows we will set aside these important caveats, and develop a theory of the compressible nematic state as an instability of the  HLR composite Fermi liquid (CFL)  state.  In Section \ref{sec:Nematic-CFL} we present the theory of the nematic composite Fermi liquid, following closely the structure and results of the theory of the nematic Fermi fluid of Oganesyan, Kivelson and Fradkin,\cite{Oganesyan-2001} and of the composite Fermi liquid of Halperin, Lee and Read\cite{Halperin-1993} Here we introduce the gauge-invariant quadrupolar interaction, and  present the basic structure of the effective action for the nematic order parameter fields. In Section \ref{sec:diagonal} we derive the parity-even part of the nematic fluctuations, and in Section \ref{sec:hall-viscosity} we derive the parity-odd component which yields the Hall viscosity using a diagrammatic approach.  
In  Section \ref{sec:Ward} we derive a set of important Ward identities which we use throughout the paper. In Section \ref{sec:dynamics} we discuss the 
effective quantum dynamics of the nematic fields and their electromagnetic response which is relevant to resonance experiments.
 In Section \ref{sec:WZ} we derive the Wen-Zee term for the CFL. Here we find that, due to the non-local nature of this term in the compressible state, not only its coefficient is not quantized (as it is in the incompressible FQH state\cite{you2014}) but its relation with the Berry phase term for the nematic fields (i.e. the effective Hall viscosity of the CFL) is not straightforward. In Section \ref{sec:hydro} we derive the geometrical response of the CFL and in particular the Hall viscosity of this compressible fluid. Here too, contrary to what happens in the FQH states, this response is not quantized and it is not universal. In Section \ref{Experiment}, we  discuss the connection of our theoretical results on the nematic CFL states with various experiments showing transport anisotropy in half-filled Landau levels. Section \ref{sec:conclusions} is devoted to our conclusions and open questions. Details of our calculations and of previous results are presented in the Appendices. In Appendix \ref{app:OKF} we give a  summary of the theory of the quantum phase transition to a nematic state in a Fermi liquid of Oganesyan and coworkers,\cite{Oganesyan-2001} and in Appendix \ref{app:HLR} we summarize  the HLR theory of the half-filled Landau level.\cite{Halperin-1993} Details of the derivation of the Berry-phase-type term for the nematic fields are given in Appendix \ref{app:details-berry} and for the Wen-Zee term for the nematic fields in Appendix \ref{app:details-WZ}. The derivation of the nematic correlators is given in Appendix \ref{app:nematic-correlators}, and the vertex correction for the nematic polarization in Appendix \ref{app:vertex}.





\section{Theory of Nematic Phase Transition of the Composite Fermi liquid}
\label{sec:Nematic-CFL}

In this Section we consider the nematic-isotropic quantum phase transition inside the CFL. We will construct the theory using the theory of the nematic transition in a Fermi liquid of Oganesyan, Kivelson and Fradkin\cite{Oganesyan-2001} (OFK), summarized in Appendix \ref{app:OKF}, and the theory of the isotropic CFL of Halperin, Lee and Read,\cite{Halperin-1993,Read-1994,Rezayi-1994,kim1994instantons} summarized in Appendix \ref{app:HLR}, as our starting points.

Our starting point is the action for electrons in a  half-filled Landau level  of HLR   with  a quadrupolar interaction 
\begin{align}
\mathcal{S} &= \int dt d^2 r  \Psi^{\dagger}(\bm{r},t) \Big[ iD_t  + \frac{\bm{D}^2}{2m} + \mu  \Big]\Psi(\bm{r}, t),\nonumber\\ 
& -\frac{1}{2} \int dt d^2 r d^2 r' ~V(\bm{r}-\bm{r}') \delta \rho(\bm{r}, t)\delta \rho(\bm{r}', t) \nonumber\\ 
&- \frac{1}{2} \int dt d^2 r d^2 r' ~F_2 (\bm{r} - \bm{r}')\text{Tr}\Big( \hat{Q}(\bm{r},t)\cdot \hat{Q}(\bm{r}',t) \Big), 
\end{align}
where  $D_{\mu} = \partial_\mu + iA_{\mu}$ is the covariant derivative and the index $\mu=t,x,y$ (not to be confused with the chemical potential  which is also denoted by $\mu$!). Here $V({\bm r}-{\bm r}')$ is the pair interaction potential, and the quadrupolar interaction $F_2$ in momentum space is represented as 
\begin{align}
F_2 (\bm{q}) = \frac{F_2}{1+\kappa \bm{q}^2}
\end{align}
Here we will use the same prescription we used in Ref.[\onlinecite{you2014}] in the context of the nematic FQH state, and we will be careful to include the gauge field $A_\mu$ in the definition of the nematic order parameter, the traceless symmetric tensor  $\hat{Q}(\bm{r})$,  
\begin{align}
\hat{Q}(\bm{r}) =\frac{1}{2m}\Psi^{\dagger}(\bm{r})
\begin{pmatrix} 
D_x^2 - D_y^2 & 2 D_x D_y \\
2D_x D_y & D_y^2 -D_x^2  
\end{pmatrix}
\Psi(\bm{r}), 
\end{align}
where $D_x$ and $D_y$ are the $x$ and $y$ components of the covariant derivative which is explicitly dependent on the gauge field $A_\mu$. Hence, the nematic order parameter couples to the electromagnetic gauge field as a quadrupole.\cite{you2014} 

\subsection{Flux attachment and nematic order}
\label{sec:attachment-nematic}

Now we proceed to attach the flux using the Chern-Simons term and follow the same strategy as in the conventional CFL theory of Appendix \ref{app:HLR} to find the following effective theory. In addition to this, we perform a Hubbard-Stratonovich transformation  to decouple the quadrupolar interaction in terms of a field ${\bm M}({\bm r},t)=(M_1({\bm r},t),M_2({\bm r},t))$    (see also Appendix \ref{app:OKF}), to find  an action of the form
\begin{widetext}
\begin{align}
\mathcal{S} &= \int dt d^2 r  \Psi^{\dagger}(\bm{r},t) \Big[ iD_t  + \frac{\bm{D}^2}{2m} + \mu+ \alpha \bm{D}^4  \Big]\Psi(\bm{r}, t) -\frac{1}{2} \int d^2r d^2r' dt \frac{1}{16 \pi^2} \delta b({\bm r},t) V({\bm r-r'}) \delta b({\bm r'},t)\nonumber\\ 
& + \int dt d^2 r \frac{\varepsilon^{\mu\nu\lambda}}{8\pi} (\delta a_{\mu}-\delta A_\mu)\partial_\nu (\delta a_\lambda-\delta A_\lambda) + \int dt d^2 r \Big[ \frac{M_{1}}{2m} \Psi^{\dagger}(\bm{r},t) (D_x^2 - D_y^2)\Psi (\bm{r},t) + \frac{M_{2}}{2m} \Psi^{\dagger}(\bm{r},t)  (2D_x D_y) \Psi (\bm{r},t)  \Big] \nonumber\\ 
& - \int dt d^2 r \frac{1}{2 F_2}\Big[ \bm{M}^2 +\kappa \left({\bm \nabla} {\bm M}\right)^2\Big] 
\label{Goal}
\end{align}
\end{widetext}
where the covariant derivative now is  $D_\mu = \partial_\mu + i \delta a_\mu$, where $\delta a_\mu$ is the fluctuating component of the Chern-Simons gauge field (See Appendix \ref{app:HLR}), and where we have used the Chern-Simons constraint to replace the density fluctuation $\delta \rho({\bm r},t)$ with the Chern-Simons  flux fluctuation $\delta b({\rm r},t)$ (as shown in Appendix \ref{app:HLR}.)

The action of Eq.\eqref{Goal} is the theory that we will analyze in this paper. This action now also contains a higher-order gradient term, $ \alpha \bm{D}^4$, in  the fermion dispersion needed to stabilize the nematic phase, i.e., making the sign of the quartic term of the free energy of the nematic order parameter positive.\cite{Oganesyan-2001} However, we will be mainly interested in the leading scaling behaviors of the various correlators in $\bm{q}$ and $\omega$ for small $q \ll k_F$ where we linearize the kinetic energy of the fermion near $k_F$ to calculate the correlators. Then the higher-order dispersion does not affect the leading scaling behaviors of the dynamic properties of the correlators. Therefore, from here and on, for the most part we we will drop the $\alpha \bm{D}^4$ term when calculating the dynamical properties of the CFL. We note that in our earlier work on the nematic fractional quantum Hall states\cite{you2014} it was necessary to include a term of order ${\bm D}^6$ to stabilize the nematic state, whereas here, in the compressible case, a term of order ${\bm D}^4$ is sufficient.

Our goal in this paper is to derive the effective theory for the external electromagnetic gauge field $\delta A_\mu$  and for the nematic order parameters  $\bm{M}$. In this section we will sketch the calculation and highlight the important features of the results. The derivations of the main results are presented in the following Sections and in the Appendices. 

We will proceed  in two stages. First we will expand the effective action  resulting from integrating out the fermions about the isotropic HLR state. The result is an effective action that depends also on the fluctuating piece of the Chern-Simons gauge field $\delta a_\mu$ (as it is done in Appendix \ref{app:HLR} for the HLR theory). After integrating out the composite fermions we obtain the following effective action
\begin{align}
\mathcal{S}_{\rm eff}[\delta a_\mu, {\bm M}, \delta A_\mu]=&-i \textrm{tr} \ln \Big[ i D_t+ \frac{{\bm D}^2}{2m}+\mu+\alpha {\bm D}^4\nonumber\\
&+\frac{M_1}{2m} \left(D_x^2-D_y^2\right)+\frac{M_2}{2m} 2D_x D_y\Big]\nonumber\\
+\int d^2r dt& \frac{1}{8\pi} \varepsilon^{\mu \nu \lambda} \left(\delta a_\mu-\delta A_\mu \right) \partial_\nu \left(\delta a_\lambda-\delta A_\lambda \right)\nonumber\\
-\frac{1}{2} \int d^2r d^2&r' dt \frac{1}{16 \pi^2} \delta b({\bm r},t) V({\bm r-r'}) \delta b({\bm r'},t)\nonumber\\
-&\int d^2r dt \frac{1}{2F_2} \left[ {\bm M}^2+\kappa \left({\bm \nabla} {\bm M}\right)^2 \right]
\label{eq:eff-action-1}
\end{align}
Since we are interested in deriving an effective field theory near the nematic (Pomeranchuk) quantum phase transition, we will expand the fermion determinant (the first two lines on the right hand side of this effective action) up to the quartic order in the nematic fields $\bm M$ (and quadratic orders in their spatial derivatives). We will also expand the effective action up to the quadratic order in the fluctuations of the Chern-Simons gauge field $\delta a_\mu$. 
Notice that the only trace of broken time reversal invariance in this effective action is in the Chern-Simons action (the third line of this effective action), and that 
the fermion determinant represents a system of composite fermions at finite density (the HLR composite Fermi liquid) coupled to the fluctuations of the 
Chern-Simons gauge fields $\delta a_\mu$ and to the nematic fluctuations  $\bm M$.

The result of this expansion is an effective action for the nematic fields $\bm M$ and an effective action for the fluctuations of the Chern-Simons gauge fields $\delta a_\mu$. The effective action has the general form
\begin{align}
\mathcal{S}[\bm{M},\delta a_\mu, \delta A_\mu]=&\mathcal{S}_n[\bm{M}]+\mathcal{S}_{\textrm{CFL}}[\delta a_\mu, A_\mu]\nonumber\\
+&\mathcal{S}_{a,M}[\delta a_\mu, \bm{M}]+\mathcal{S}_{a,M,a}[\delta a_\mu, \bm{M}]
\label{eq:Stotal}
\end{align}
The first two terms of the right hand side are the expected terms for the effective action for nematic fields alone, $S_{n}[\bm{M}]$,   identical to the  result of OKF for effective nematic theory  (shown in Eq.\eqref{eq:OKF} of Appendix \ref{app:OKF}, and subsequent equations), and terms for the Chern-Simons gauge fields alone, $\mathcal{S}_{\textrm{CFL}}$ identical to the HLR result (given explicitly in Eq.\eqref{effectiveCFL} of Appendix \ref{app:HLR}.). Thus, the nematic order parameter fields $\bm{M}$ condense at the Pomeranchuk instability and become overdamped (with dynamical critical exponent $z=3$.) Likewise, the Chern-Simons gauge fields are  overdamped and also have $z=3$ dynamic critical exponent. In both theories, the fermionic quasiparticles are destroyed by these overdamped fluctuations.

The  physics of the  nematic composite Fermi fluid originates in the last two terms of the effective action of Eq.\eqref{eq:Stotal}. Although the nematic order parameters are charge-neutral, they still  couple to the gauge fields but as a quadrupole. This coupling leads to two new terms in the effective action, not present either in the theory of the nematic Fermi fluid, or in the  theory of the composite Fermi liquid. The effective action $\mathcal{S}_{a,M}$ in Eq.\eqref{eq:Stotal} has the form (see Eq.\eqref{eq:SaM} of Appendix \ref{app:coupling-gauge} for details)
\begin{equation}
\mathcal{S}_{a,M}[\bm{M},\delta a_\mu]=-\frac{1}{2}\int_{\bm{q},\omega} M_i (\bm{q},\omega) T_{i \nu}(\bm{q},\omega) \delta a_{\nu}(\bm{q},\omega)
\label{eq:SaM-body}
\end{equation}
This term is a mixed bilinear form in the nematic field $\bm{M}$ and in the Chern-Simons gauge field $\delta a_\mu$, and represents the quadrupolar coupling. 
 
 The other new term, represented by the effective action $\mathcal{S}_{a,M,a}$ of Eq.\eqref{eq:Stotal}, has the form of (see Eq.\eqref{eq:SaMa} of Appendix \ref{app:coupling-gauge} for details)
  \begin{equation}
\mathcal{S}_{a,M,a}[\delta a_\mu,\bm{M}]=-\frac{1}{2}\int_{\bm{q},\omega}\delta a_{\mu}(\bm{q},\omega) V_{\mu \nu}[\bm{M}] \delta a_{\nu}(\bm{-q},-\omega),
\label{eq:SaMa-body}
\end{equation}
which represents the parity-even coupling Maxwell-type terms of the Chern-Simons gauge fields to the local fluctuations of the nematic order parameters. In this last term, the nematic fields $\bm{M}$ couple to the gauge fields as a fluctuating spatial metric.

The couplings between the Chern-Simons gauge fields and the nematic fields in Eq.\eqref{eq:Stotal} imply that these fields mix. This has important consequences for the effective dynamics of the nematic fields. This is found in our third and last step in which we now integrate out the fluctuations of the Chern-Simons gauge fields. Since the effective action of the Chern-Simons gauge fields has a Chern-Simons term which is odd under parity and time-reversal, this step leads to parity-odd terms in the effective action of the nematic fields. Also, from the form of the coupling to the external electromagnetic fields, we will now obtain an effective action for these probe fields (the same as in the HLR theory) plus their quadrupolar coupling to the nematic fields. This last effective coupling leads to the signatures of the nematic fluctuations (and order) in the current correlation functions. 

In Ref.[\onlinecite{you2014}] we presented a theory of a nematic FQH state based on a Chern-Simons gauge theory of flux attachment. An important feature of that theory is that already at the mean filed level (i.e. the average field approximation) the effective action of the nematic fields has a Berry phase term, originating from the broken time reversal invariance, which dictates the quantum dynamics. The actual coefficient of this term can be exactly obtained at the level of mean field theory, and further gauge fluctuation correction does not modify the result. This coefficient is part of the actual, universal, value of the Hall viscosity. However, in the case of a nematic composite Fermi liquid the situation is quite different since at the mean field level (where the gauge fluctuation is ignored) the gapless composite fermions do not see directly a broken time reversal invariance which is encoded  in the Chern-Simons action of the gauge field fluctuations. We will see below that the fluctuations of the Chern-Simons gauge fields will induce a Berry phase type term for the nematic fields although this term will  be non-local and its coupling constant is unquantized (and non-universal). The same holds for the Wen-Zee term and  the Hall viscosity.

\subsection{Effective field theory of nematic fluctuations}
\label{sec:effective-nematicity}

Before  proceeding further, we first relate the various parameters appearing in Eq.\eqref{Goal}, i.e., the Fermi momentum and the effective mass of composite fermions, to the natural scales in a landau level: the magnetic length $l_0$ and effective interaction strength between electrons.
In the Landau level, the density of electron is naturally related with the magnetic length. On the other hand, the density is related with the Fermi momentum $k_F$, resulting in the standard relations
\begin{equation}
k_F=\sqrt{4\pi \rho}=\frac{1}{l_0},
\label{eq:unit}
\end{equation}
where we use units in which $\hbar/ec=1$.
On the other hand, the mass of the composite fermion is renormalized by the interactions between electrons. In the limit of a large magnetic field the effective mass is expected to be determined by the scale of electron-electron interactions alone, as shown by the work of  Halperin, Lee and Read.\cite{Halperin-1993}  We can estimate the mass of composite fermions in terms of the interactions via the dimensional analysis (see Appendix \ref{app:HLR}). Hence we obtain an estimate of the effective mass $m$ of the composite fermions in terms of the density-density interaction $V(\bm{q})$ as
\begin{equation}
m=\frac{C}{V(\bm{0})}, 
\label{eq:unit2}
\end{equation}
where  $C$ is a numerical constant. In the case of Coulomb interactions, HLR showed that  the energy scale is the Coulomb energy at the scale of the magnetic length. Here we are working with a model with short range interactions and hence we will set the effective mass to be given by Eq.\eqref{eq:unit2}. Further, the high-frequency cutoff $\bar \Lambda$ is naturally  the Fermi energy, i.e., 
\begin{equation}
{\bar \Lambda}=E_F^{\rm CFL}=\ell_0^{-2}/m.
\label{eq:unit3}
\end{equation}
 With these relations at hand, we can express all quantities in terms of the magnetic length $l_0$ and of the interaction $1/V(\bm{0})$.

We will work close to the Pomeranchuk quantum phase transition to the nematic state, where the distance to the nematic quantum critical point (the Pomeranchuk instability) is parametrized by 
\begin{equation}
\delta=\frac{V(0)}{4\pi l_0^4}+\frac{1}{ 2 F_2}.
\label{eq:Pomeranchuk}
\end{equation}
We will keep terms in the effective action up to (and including) quartic order in the nematic fields. $N_F$ is the density of states at the Fermi surface. Here, as in the HLR work on the composite Fermi liquid, we will keep only terms in the effective action which are quadratic in the fluctuations of the gauge fields $\delta a_\mu$. This amounts to working in the random phase approximation (RPA). Higher order terms are (presumably) unimportant and we will neglect them. The effective action derived below is effectively a loop expansion in the fluctuations of the gauge fields. Although in most cases we will need to consider diagrams with up to two internal gauge field propagators, in the case of the Wen-Zee term the leading non-vanishing three has three internal gauge field propagators.

Before presenting the details of our theory, we first summarize the main results. Here we show that the effective action for the nematic fields $\bm M$ is
\begin{widetext}
\begin{align}
\mathcal{S}_{\rm eff}[\bm{M},\delta A_\mu]=&\int_{\bm{p},\Omega} M_i(\bm{p},\Omega) \mathcal{L}_{ij}(\bm{p},\omega) M_j(-\bm{p},-\Omega)
- \int d^2r dt  \lambda \bm{M}^4  \nonumber\\
&+ \int_{{\bm p},\Omega} \chi \; i \Omega\  \varepsilon_{ij} M_i(\bm{p},\Omega)  M_j(-\bm{p},-\Omega)
+\int_{\bm{p},\Omega} \beta  \epsilon^{\mu \nu \rho}
\omega_{\mu} (\bm{p},\Omega) \Big(\partial_{\nu} \delta A_{\rho}\Big)(-\bm{p},-\Omega)\nonumber\\ 
&-\int_{\bm{p},\Omega} \frac{1}{2}\delta A _{\mu}(\bm{p},\Omega) \left(V_{\mu \nu}(\bm{p},\Omega)[\bm{M}] + K_{\mu\nu} (\bm{p},\Omega)\right) \delta A_{\nu} (-\bm{p},-\Omega), 
\label{EffectiveTheory}
\end{align}
where\cite{Oganesyan-2001} $\lambda=(1/l_0^2)^{3}(2\alpha)/5$, with $\alpha$ being the coefficient of the quartic term in the single-particle dispersion (see Appendix \ref{app:OKF}), and the coefficient $\beta$ is given by $\beta=\frac{2}{3\pi \sqrt{3}}$. In Eq.\eqref{EffectiveTheory} $\mathcal{L}_{ij}({\bm p},\omega)$ is the inverse propagator of the nematic fields  
\begin{equation}
\mathcal{L}_{ij}(\bm{p},\Omega)=
\begin{pmatrix} 
-\frac{\kappa}{2F_2} p^2-\delta -\cos^2(2 \theta_p)\frac{1}{2\pi l_0^3} \left(\frac{i\Omega}{p}\right) &  \sin(2 \theta_p)\cos(2 \theta_p)\frac{1}{2\pi l_0^3}\left(\frac{i\Omega}{p}\right)
 \\ 
\sin(2 \theta_p)\cos(2 \theta_p)\frac{1}{2\pi l_0^3} \left(\frac{i\Omega}{p}\right)& -\frac{\kappa}{2F_2} p^2-\delta -\sin^2(2 \theta_p)\frac{1}{2\pi l_0^3} \left(\frac{i\Omega}{p}\right)
\end{pmatrix}.
\label{eq:calLij}
\end{equation}
where, as in Eq.\eqref{eq:Pomeranchuk}, $\delta$ denotes the distance to the quantum critical point (i.e. the Pomeranchuk instability), and $\theta_p$ is teh angle of the momentum $\bm p$ with the $x$ axis. The result of Eq.\eqref{eq:calLij} was first derived by Oganesyan {\it et al.}.\cite{Oganesyan-2001}

The second term in Eq.\eqref{EffectiveTheory} is a Berry phase term and its (non-universal) coefficient $\chi=\frac{2}{3\pi l_0^2} $ is the Hall viscosity of the CFL (see Section \ref{sec:hall-viscosity}). Finally, the tensor $V_{\mu \nu}[\bm M]$  represents the parity-even  coupling between the Maxwell terms of the electromagnetic gauge fields and the nematic fields (and couple as a metric fluctuation) is given by
\begin{equation}
V_{\mu \nu}(\bm{p}, \omega)[\bm{M}]=\frac{1}{2\pi}  K_0(\bm{p}, \omega)
\begin{pmatrix} 
\frac{M_1}{2}(p_x^2-p_y^2) & (M_1 p_x+M_2 p_y)\omega  & (-M_2 p_y+M_2 p_x)\omega  \\
(M_1 p_x+M_2 p_y)\omega & M_1\omega^2  & M_2\omega^2 \\
(-M_2 p_y+M_2 p_x)\omega & M_2 \omega^2 & -M_1\omega^2 
\end{pmatrix}.
\label{Propagator}
\end{equation}
\end{widetext}
 The function $K_0(\bm{p}, \omega)$ is given in Appendix \ref{app:HLR}, and $K_{\mu \nu}({\bm p},\omega)$ is the polarization tensor of the electromagnetic field in the HLR theory of the CFL.

We first remark that, due to the Landau damping terms in the inverse propagator $\mathcal{L}_{ij}({\bm p},\Omega)$,  the nematic phase transition of the compressible half-filled Landau level has a dynamical critical exponent $z=3$. However, the effective action of Eq.\eqref{EffectiveTheory} has also a Berry-phase-type term for the nematic order parameters induced by Chern-Simons gauge fluctuation. Although this term is formally subleading to the Landau damping term, it is kept since it is the leading parity-odd contribution to the nematic fields.

By symmetry, the nematic order parameter acts as a locally-fluctuating dynamical metric to electrons and to composite fermions and modifies the local frames. In the nematic phase, where the order parameters have a non-vanishing expectation value,  this coupling leads to an anisotropic  electromagnetic response. These effects are encoded through the term $V_{\mu\nu}$ of the polarization tensor for the external probe electromagnetic gauge field $\delta A_\mu$. The isotropic part $K_{\mu\nu}$ of the polarization tensor was calculated in Ref.[\onlinecite{Halperin-1993}] (see Eq.\eqref{polEM}). Here $\omega_\mu$ plays the role the ``spin connection" of the ``dynamical metric" defined by the nematic order parameter fields, and can be explicitly written out in terms of the nematic order parameter fields as
\begin{align}
&\omega_{0}=\epsilon^{ij} M_i \partial_0 M_j \\\nonumber
&\omega_{x}=\epsilon^{ij} M_i \partial_x M_j-(\partial_x M_2-\partial_y M_1) \\\nonumber
&\omega_{y}= \epsilon^{ij} M_i \partial_y M_j +(\partial_x M_1+\partial_y M_2)
\end{align}
Due to the Wen-Zee-like term, i.e., the term $\sim \epsilon^{\mu\nu\lambda}\omega_\mu \partial_\nu \delta A_\lambda$ in the action, the disclination of the nematic order parameters inside the nematic phase minimally couples with the electromagnetic gauge field and carries the (non-quantized) electric charge, which, in this compressible state, will be eventually screened by the gapless electrons. 

The effective theory is both gauge invariant  and rotationally invariant. First of all, the inverse propagator $\mathcal{L}_{ij}$ in Eq.\eqref{EffectiveTheory} and Eq.\eqref{Propagator} of the nematic order parameter is constructed in the way that it is apparently rotationally symmetric. We will come back to this later. Secondly, the action is also gauge invariant. The Wen-Zee-like term is apparently gauge invariant because it involves the field strength $F_{\nu\rho} = \partial_\nu \delta A_\rho -\partial_\rho \delta A_\nu$ explicitly. For the full polarization tensor $V_{\mu\nu}+K_{\mu\nu}$ for the external electromagnetic gauge field $\delta A_\mu$, it is clear that the gauge invariance is respected, since $\partial_\nu V_{\nu\lambda} = 0$ and $\partial_\nu K_{\nu\lambda} =0$, which implies  gauge invariance. 


We present the detailed calculation for the main results in the following sections.

\section{Parity-even Components of the nematic fluctuations}
\label{sec:diagonal}

\begin{figure} [hbt] 
\begin{center}  
\includegraphics[width=0.4\textwidth]{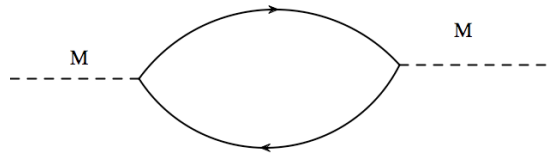}  
\caption{Nematic correlator: the full lines represent the composite fermion propagator, and the broken lines represent the nematic order parameters.}
\label{NCorrelator}
\end{center}  
\end{figure} 

The propagator of the nematic order parameters in the Fermi liquid state is (here $\mathcal{T}$ denotes time ordering and $i,j=1,2$)
\begin{equation}
i \langle \mathcal{T} M_i M_j\rangle_0 (\bm{p},\omega)\equiv \mathcal{L}^{-1}_{ij}(\bm{p},\omega)
\end{equation}
The  diagonal component  $\mathcal{L}_{11}$ (represented diagrammatically in Fig.\ref{NCorrelator}) 
has the  explicit form 
\begin{align}
&\mathcal{L}_{11}(\bm{p},\Omega) =-i \int_{\bm{k}, \omega}~ g (\bm{k+p},\Omega+\omega)  g (\bm{k},\omega) \left(\frac{k_x^2-k_y^2}{2m}\right)^2 \nonumber\\
&=\frac{k_F^4}{4\pi m}-\cos^2(2 \theta_p)\frac{k_F^3}{2\pi} \left(\frac{i\Omega}{p}\right)+o(\Omega/p), \nonumber\\
&=\frac{ V(0)}{4\pi l_0^4}-\cos^2(2 \theta_p)\frac{1}{2\pi l_0^3} \left(\frac{i\Omega}{p}\right)+o(\Omega/p)
\end{align}
where 
\begin{equation}
g(\bm{k}, \omega) = \frac{1}{\omega - \frac{\bm{k}^2 - k_F^2}{2m} + i \eta  \;  \sign(\omega)}
\end{equation}
 is the time-ordered free composite fermion propagator. 
 In the same way, we can calculate the other components of $\mathcal{L}_{ij}(\bm{p},\omega)$ and replace $k_F,m$ in terms of $l_0,V(0)$,
 \begin{align}
\mathcal{L}_{12}(\bm{p},\Omega)=&\cos(2 \theta_p)\sin(2 \theta_p) \frac{1}{2\pi l_0^3} \left(\frac{i\Omega}{p}\right), \nonumber\\
\mathcal{L}_{22} (\bm{p},\Omega)=&\frac{V(0)}{4\pi l_0^4}-\sin^2(2 \theta_p)\frac{1}{2\pi l_0^3} \left(\frac{i\Omega}{p}\right), 
\end{align}
which agree with the results of  OKF.\cite{Oganesyan-2001}

Now we can include the gauge fluctuations. This is included through the loop expansion in the gauge fields, and we calculate only the one-loop corrections.
In Appendix  \ref{app:nematic-correlators} and Appendix \ref{app:vertex} we show that these corrections  are subleading to the leading term, and thus do not change the dynamic scaling behavior of the critical theory of the nematic-isotropic phase transition.

\section{Parity-odd Components of the nematic fluctuations: the Hall viscosity}
\label{sec:hall-viscosity}

In our earlier work,\cite{you2013,you2014} we  investigated the isotropic-nematic phase transition in FQH states and Chern insulators. We  concluded that the theory describing the  phase transition to the anisotropic state in such chiral topological phases always contains a Berry phase term for the nematic order parameters, which is odd under time reversal and parity. In the nematic FQH states the coefficient of the Berry phase term is related (but not equal to) with a dissipationless Hall viscosity.\cite{you2014, Cho-2014} In Ref.[\onlinecite{you2014}] we concluded that the nematic fluctuation, regarded as a dynamical metric, only couples with the stress tensor of the composite fermion while the background metric also appears in the covariant derivative as the spin connection of the composite particle. Accordingly, the Berry phase of the nematic order parameter \textit{is} the odd Hall viscosity of the mean-field state of the composite fermion theory, not of the electron fluid. Hence, in the incompressible FQH states, the Berry phase term is equivalent to the Hall viscosity of the composite fermion filling up integer number of the effective Landau levels.\cite{Cho-2014, you2014}

For the compressible half-filled Landau levels, the mean field state of the composite fermion without the gauge fluctuation \textit{is} a CFL with well-defined Fermi surface. Furthermore, the state is time-reversal even so we do not expect any Berry phase term for the nematic order parameters to emerge. However, once we include the dynamics of the gauge fluctuation $\delta a_\mu$ and go beyond the mean-field theory, the time-reversal symmetry is explicitly broken due to the Chern-Simons term for $\delta a_\mu$ and thus the Hall viscosity is expected to arise from the fluctuations. This is to be expected since, for the same reasons,  the Hall conductivity of the CFL in the HLR theory also comes from gauge fluctuations. 

To study the contribution from the gauge fluctuation to the Berry phase term, we formally integrate out the fermions at  one-loop order, and rewrite the theory in terms of the fluctuating gauge field $\delta a_\mu$ and the nematic order parameter. Throughout this section we use the polarization functions of the compressible fermions $\Pi_{\mu \nu}$, $\Pi_0$ and $\Pi_2$, defined in Eqs.\eqref{eq:Pimunu}, \eqref{eq:Pi0} and \eqref{eq:Pi2}, respectively, given explicitly in Appendix \ref{app:coupling-gauge}. We first consider the linear coupling between the nematic field and the gauge field at the mean field level,
\begin{equation}
\mathcal{S}_{a,M}=-\int_{\bm{q},\omega}~\frac{1}{2} M_i(\bm{q},\omega) T_{i \nu}(\bm{q},\omega) \delta a_{\nu}(\bm{-q},-\omega),
\end{equation}
where $T_{i\nu}(\bm{q},\omega)$ is the $2 \times 3$ matrix (with $i=1,2$ and $\nu=t, x, y$)
\begin{equation}
 T_{i\nu}(\bm{q},\omega)=\frac{ m}{2\pi}\Pi_2(\bm{q},\omega)
\begin{pmatrix} 
q_x^2-q_y^2 & q_x\omega & -q_y\omega\\
-2q_xq_y & -\omega q_y  & -q_x \omega
\end{pmatrix}.
\end{equation}
is the effective vertex.
This coupling can be regarded as the symmetric part of the Wen-Zee-like term which we will discuss later. 

Beyond this, there is another term involving the fluctuating nematic order parameter and Chern-Simons gauge fields. 
\begin{equation}
\mathcal{S}_{a,M,a}=-\frac{1}{2}\int_{\bm{q},\omega}\delta a_{\mu}(\bm{q},\omega) \mathcal{V}_{\mu \nu}(\bm{q},\omega)\delta a_{\nu}(\bm{-q},-\omega),
\end{equation}
where
\begin{widetext}
\begin{equation}\mathcal{V}_{\mu \nu}(\bm{q},\omega)=\frac{\Pi_0(\bm{q},\omega)}{2\pi} \begin{pmatrix} 
\frac{M_1}{2}(q_x^2-q_y^2) & (M_1 q_x+M_2 q_y)\omega  & (-M_2 q_y+M_2 q_x)\omega  \\
(M_1 q_x+M_2 q_y)\omega & M_1\omega^2  & M_2\omega^2 \\
(-M_2 q_y+M_2 q_x)\omega & M_2 \omega^2 & -M_1\omega^2 
\end{pmatrix}.
\end{equation}
\end{widetext}
This term is the coupling to the dynamical metric or nematic order parameter of the Maxwell term for the fluctuating gauge field $\delta a_\mu$. As the nematic order parameter can be considered as a dynamical metric that modifies the local metric of the composite fermion, the gauge boson coupled to the fermion is also modified accordingly by the nematic order parameter.

Together with the original isotropic HLR result for the action of the fluctuating Chern-Simons field $\delta a_\mu$
\begin{align}
\mathcal{S}_{a}= -\frac{1}{2}\int_{\bm{q}, \omega} \delta a_{\mu}(\bm{q}, \omega) \Pi_{\mu \nu}(\bm{q},\omega) \delta a_{\nu}(-\bm{q},-\omega),
\end{align} 
we can obtain the Berry phase term of the nematic order parameter by integrating out the fluctuating gauge fields and performing the loop expansions. In the following, we fix the gauge $a_0=0$ to facilitate the calculation.

The effective action   $\mathcal{S}= \mathcal{S}_a + \mathcal{S}_{a,M,a}$ of the gauge field $\delta a_\mu$, defined by Eq.\eqref{eq:SaM-body} and Eq.\eqref{eq:SaMa-body}, coupled with the nematic order parameter $\bm M$,  is 
\begin{widetext}
\begin{align}
&\mathcal{S}=-\frac{1}{2}\int_{\bm{q},\omega,\bm{k},\Omega} \delta a_{i}(\bm{q},\omega) \Big[ \Pi_{ij} +(t^1_{ij} \omega^2 \Pi_0)  M_1 + (t^2_{ij} \omega^2 \Pi_0 )M_2 \Big]\delta a_{j} (\bm{-q},-\omega)
\end{align}
where the $2\times 2$ matrices $t^1_{ij}$ and $t^2_{ij}$ denote the Pauli matrices $\sigma_3$ and $\sigma_1$, respectively, and $\bm M$ are the nematic fields.
By integrating out the fluctuations of the gauge fields, and performing the loop expansion, we obtain leading corrections to the   (time-ordered) correlators of the nematic order parameter $\bm M$ (shown in the Feynman diagram of Fig. 2)
\begin{align}
&\langle  \delta M_i \delta  M_j\rangle(\Omega,\bm{p})= 
i  \int_{\bm{q},\omega} \textrm{tr}\left[ \Pi_{ij}^{-1}(\omega,\bm{q}) \; \omega^2 \Pi_0 t^{i}\;  \Pi_{ij}^{-1}(\omega+\Omega,\bm{q+p}) \; \omega^2 \Pi_0 t^{j} \right]\nonumber\\
&=i\frac{\Omega}{2} \epsilon^{ij}~\int_{\bm{q},\omega}~ \frac{\Pi_2 \bm{q}^2 }{\Pi_2 \bm{q}^2+\bm{q}^2/4} \frac{ \sqrt{{\bar \rho}}}{\Pi_2 \bm{q}^2+\bm{q}^2/4}+....
\end{align}
\end{widetext}
where we have approximate the form of the polarization tensor $\Pi_{ij}(\omega,\bm q)$ valid for low frequency and momentum with $\omega \ll |\bm q| v_F$ (see Appendix \ref{app:HLR}).

\begin{figure}[hbt] 
\begin{center}  
\includegraphics[width=0.4\textwidth]{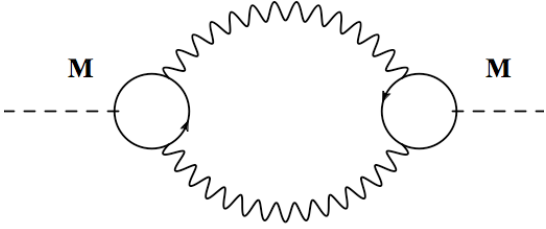}  
\caption{Berry phase term. Here the  bubble is the fermion loop corrected by the fluctuating gauge field.}
\end{center}  
\label{fig:berry-nematic}
\end{figure}  

Hence, the anti-symmetric (and hence off-diagonal) part of the nematic correlator $\mathcal{L}^{-1}_{ij}$ gives a Berry phase type term in the effective action for the nematic fields
\begin{equation}
\mathcal{S}^{M}_{ij}= \int_{\bm{p},\Omega}\chi \epsilon^{ij} (i \Omega) M_i(\bm{p},\Omega) M_j(-\bm{p},-\Omega),
\label{eq:berry-cfl}
\end{equation}
where 
\begin{equation}
\chi= \frac{2}{3\pi}  \bar{\Lambda} m=\frac{2}{3\pi l_0^2} 
\end{equation}
with  $\bar \Lambda$  the high-frequency cutoff of the CFL and $m$ is the effective mass of the composite fermions. Using the HLR results\cite{Halperin-1993} (see Appendix \ref{app:HLR}), since the UV cutoff ${\bar \Lambda}=E_F^{\rm CFL}$ (the Fermi energy of the CFL), and the value of the effective mass $m$ (which in the HLR theory is argued to depend only on the scale of the Coulomb interaction) one finds that  $m{\bar \Lambda}=\ell_0^{-2}$, and hence that the Hall viscosity  seemingly depends only on the particle density. This  
 value of the Berry phase $\chi$  is one of our main results in this paper. It plays a key role in the effective dynamics of the nematic order parameters. However, we should caution that, contrary to the case of the FQH states, this  value of the Hall viscosity is not protected in  the CFL, and should be regarded as an estimate.

Our result indicates that the gauge field fluctuations generate a Berry phase term which will in turn contribute to the Hall viscosity of the compressible half-filled Landau levels. The way in which the Berry phase term is generated in this case is different from that of the incompressible FQH states. In the incompressible FQH states, the Berry phase of nematic order parameters  is already present  at the mean field level of the composite fermion which forms an integer quantum Hall state, due to the explicitly broke  the time-reversal symmetry of the composite fermions in the effective Landau level.\cite{Cho-2014,you2014} Furthermore, the Chern-Simons gauge fields in the FQH state are gapped and do not affect the value of the Berry phase term in the low energy regime.
 In contrast, for compressible half-filled Landau levels, the mean field state is a Fermi liquid alone which seemingly respects time-reversal symmetry, which is broken by the Chern-Simons term for the gauge fields. Their fluctuations can affect the low-energy dynamics of the nematic order parameters and induce the non-zero Berry phase term for the nematic order parameters. In particular, while in the incompressible FQH states the coefficient of the Berry phase term has a universal relation with the composite fermion density, in the case of the compressible state this ``Hall viscosity'' is non-universal as it depends explicitly on the UV energy cutoff $\bar \Lambda$ of the compressible composite Fermi fluid.

\section{Ward Identities}
\label{sec:Ward}

We now present another way to compute and check the Berry phase term, Eq.\eqref{eq:berry-cfl}. Here we will use  the Ward identity between the current operators $\bm J$ and the stress tensor $T_{ij}$ (and the linear momentum density $T_{0j}$). We start with the explicit expressions for these operators in the CFL, 
\begin{align}
&J_i=\frac{1}{2m}[\Psi^{\dagger}(\bm{r},t) (\partial_{i}+i\delta a_i) \Psi-((\partial_{i}+i\delta a_i) \Psi^{\dagger}) \Psi(\bm{r},t)],\nonumber\\
&T_{ij}=\frac{1}{2m}\Psi^{\dagger} (\bm{r},t) (\partial_{i}+i\delta a_i)(\partial_{j}+i\delta a_j) \Psi(\bm{r},t),\nonumber\\
&T_{0j}=m J_j. 
\end{align}
which are manifestly gauge invariant. 

The conservation law of the energy-momentum tensor (i.e. local conservation of energy and momentum), in the presence of the Chern-Simons gauge field fluctuations, implies that
\begin{align}
\partial_{\mu}T_{\mu i}=\delta b ~\epsilon^{ij} J_j,
\label{eq:conservation-T}
\end{align}
where $\delta b=\varepsilon_{ij} \partial_i \delta a_j $ is the fluctuating flux of the gauge field $\delta a_\mu$.
Upon expanding out  in components the expression of Eq.\eqref{eq:conservation-T}, we have
\begin{align}
\partial_0 m J_x+\partial_{x}T_{xx}+\partial_{y}T_{yx}=\delta b  J_y, \nonumber\\
\partial_0 m J_y+\partial_{x}T_{xy}+\partial_{y}T_{yy}=-\delta b  J_x. 
\end{align}
In what follows we will use the notation $T_1=T_{xx}-T_{yy}$, and  $T_2=T_{xy}+T_{yx}$. Focusing only on the anti-symmetric response, we find the following relation between the correlation functions of current, stress tensor and density operators,
\begin{align}
&\partial^2_{t_1} m^2 \langle J_x (\bm{r}_1,t_1)J_y (\bm{r}_2,t_2)\rangle +\partial^2_{\bm{r}_1}  \langle T_1(\bm{r}_1,t_1)T_2(\bm{r}_2,t_2)\rangle \nonumber\\
&=4\langle\rho(\bm{r}_1,t_1)J_x (\bm{r}_1,t_1)\rho(\bm{r}_2,t_2)J_y(\bm{r}_2,t_2)\rangle.
\label{eq:conservation}
\end{align}
Hence the Berry phase term, i.e. the Hall viscosity determined by the parity-odd correlation function of the stress tensor,\cite{Read2011,Bradlyn-2012,Gromov2014,Cho-2014,Gromov-2015,you2014,Bradlyn-2015}  is related with the  correlation function of the composite operators of densities and currents (shown on on the right side of Eq.\eqref{eq:conservation}). In momentum and frequency space (and in the low frequency limit, $\Omega \to 0$) this results takes the form
\begin{align}
\bm{p}^2 \langle T_1T_2 \rangle(\bm{p},\Omega)
&=\int_ {\bm{q} ,\omega}~ [4\langle \rho  \rho \rangle(\bm{p-q},\omega+\Omega)\langle J_x J_y \rangle(\bm{q},\omega)\nonumber\\
&+ 4\langle \rho  J_y \rangle(\bm{p-q},\omega+\Omega) \langle J_x \rho\rangle(\bm{q},\omega)]
\end{align}
 
The density and current correlators, $\langle \rho  \rho\rangle$ and $ \langle J_x J_y \rangle$, are given by the polarization tensor of the external electromagnetic gauge field beyond the mean field level which include the fluctuations of the Chern-Simons gauge field. 
\begin{align}
&\langle \rho  \rho \rangle(\omega,\bm{q})=  \frac{1}{4(2\pi)} \frac{\Pi_0 q^2}{D(\omega, \bm{q})}  \nonumber\\
&\langle J_x J_y \rangle(\omega,\bm{q})= \frac{i\omega}{2(2\pi)}, \nonumber\\
&\langle \rho  J_y \rangle(\omega,\bm{q})=-\frac{1}{4(2\pi)}\frac{\Pi_0 q_y \omega}{D(\omega, \bm{q})}-\frac{i q_x}{2(2\pi)},\nonumber\\
&\langle J_x \rho \rangle(\omega,\bm{q})=-\frac{1}{4(2\pi)}\frac {\Pi_0 q_x \omega}{D(\omega, \bm{q})}-\frac{i q_y}{2(2\pi)}, \nonumber\\
&D(\omega,\bm{q})=\frac{1}{2\pi}\Big(\Pi_0^2 \omega^2-\frac{1}{4}-\Pi_0 q^2 (\Pi_2+\frac{V(q)}{16\pi^2})\Big). 
\end{align}
Using  these relations we obtain a result  consistent with  the previous approach,
\begin{align}
&\langle T_1 T_2 \rangle (\Omega, \bm{p})=\partial_{p^2} 4 \langle \rho  J_x  \rho  J_y \rangle (\Omega, \bm{p})\nonumber\\
&=i\Omega \frac{2}{3\pi} m  \bar{\Lambda}=i\Omega \frac{2}{3\pi l_0^2}  
\end{align}

\section{Effective Dynamics and Susceptibility of the Nematic Order}
\label{sec:dynamics}

With the results of the calculation of the Berry phase term and Landau damping terms for the nematic order parameters at hand, we can  proceed to derive the  expression for the nematic correlators including both contributions. 
The effective theory for the nematic order parameter field $\bm M$ close to the nematic  transition  is  
\begin{align}
\mathcal{S}[\bm{M}]=&\int_{\bm{p},\Omega} M_i(\bm{p},\Omega) \mathcal{L}_{ij} M_j(-\bm{p},-\Omega)-\int d^3x \lambda \bm{M}^4
\end{align}
in which the correlator $\mathcal{L}_{ij}(\bm{p},\omega)$ is given by the sum of the two contributions, which yields the result
\begin{widetext}
\begin{align}
&\mathcal{L}_{ij}=\begin{pmatrix} 
-\frac{\kappa}{2F_2} p^2-\delta -\cos^2(2 \theta_p)\frac{1}{2\pi l_0^3} (\frac{i\Omega}{p}) &  i\Omega( \frac{2}{3\pi l_0^2}  )+ \sin(2 \theta_p)\cos(2 \theta_p)\frac{1}{2\pi l_0^3} (\frac{i\Omega}{p})  \\ 
 -i\Omega (\frac{2}{3\pi l_0^2} )+\sin(2 \theta_p)\cos(2 \theta_p)\frac{1}{2\pi l_0^3} (\frac{i\Omega}{p})& -\frac{\kappa}{2F_2}  p^2-\delta -\sin^2(2 \theta_p)\frac{1}{2\pi l_0^3} (\frac{i\Omega}{p})
\end{pmatrix}, 
\label{eq:Lij-CFL}
\end{align}
\end{widetext}
where $\delta$ is the distance to the Pomeranchuk instability of  Eq.\eqref{eq:Pomeranchuk}.
As in the case of the nematic Fermi fluid, $\mathcal{L}_{ij}(\bm{p},\omega)$ is nothing but the inverse susceptibility of the nematic order parameters, and zeros of the determinant of $\mathcal{L}_{ij}(\bm{p},\omega)$  yield the dispersion relation of the nematic collective modes of the CFL. Clearly the nematic susceptibility is finite for $\delta>0$ and diverges as $\delta \to 0$, as expected at a continuous quantum phase transition.

Except for the Hall viscosity term,  which originates from the gauge fluctuations which explicitly break the time-reversal symmetry, Eq.\eqref{eq:Lij-CFL} is almost the same as the result for the nematic correlator of OKF.\cite{Oganesyan-2001}  The Berry phase term, which makes the effective theory time-reversal odd, mixes  the transverse and longitudinal modes of the nematic order parameters. However, if we focus only on the parameter regime where $q\ll k_F$ and $\omega \ll v_Fq $, then the off-diagonal terms are sub-dominant and can be ignored. Hence the overdamped critical mode is not affected by the  gauge fluctuations. This leads to the conclusion that the criticality of the isotropic-anisotropic phase transition of half-filled LL still exhibits $z=3$ critical dynamical exponent. 

\subsection{Nematic Susceptibility}

From the effective theory Eq.\eqref{eq:Lij-CFL}, we can read-off the dynamic nematic susceptibility $\chi^M$, i.e. the nematic propagator. 
Inside the isotropic phase, $\delta>0$, the susceptibility is given by
\begin{widetext}
\begin{equation}
\chi^M_{ij}=
\mathcal{L}^{-1}_{ij}
=\frac{1}{C(\Omega,{\bm p})}
\begin{pmatrix} 
-\frac{\kappa}{2F_2} p^2-\delta -\sin^2(2 \theta_p)\frac{1}{2\pi l_0^3} (\frac{i\Omega}{p}) &  i\Omega( \frac{2}{3\pi l_0^2} )- \sin(2 \theta_p)\cos(2 \theta_p)\frac{1}{2\pi l_0^3} (\frac{i\Omega}{p})\\
 -i\Omega (\frac{2}{3\pi l_0^2})-\sin(2 \theta_p)\cos(2 \theta_p)\frac{1}{2\pi l_0^3} (\frac{i\Omega}{p})& -\frac{\kappa}{2F_2} p^2-\delta -\cos^2(2 \theta_p)\frac{1}{2\pi l_0^3} (\frac{i\Omega}{p})
\end{pmatrix}
\end{equation}
where
\begin{equation}
C(\Omega,\bm{p})=(\frac{\kappa}{2F_2} p^2+\delta)^2+(\frac{\kappa}{2F_2} p^2+\delta)\frac{1}{2\pi l_0^3} (\frac{i\Omega}{p})-\Omega^2 (\frac{2}{3\pi l_0^2}  )^2 \approx\delta^2+\delta \frac{1}{2\pi l_0^3} \frac{i\Omega}{p}.
\end{equation}
\end{widetext}
The nematic susceptibility is finite in the isotropic phase where $ \frac{1}{2\pi l_0^3} \frac{i\Omega}{p}\ll1$. As we approach the quantum critical point, $\delta \rightarrow 0$, the nematic susceptibility diverges, as expected.
In the nematic phase, we assume the the nematic order is in the $M_1$ direction ($M_1=\sqrt{\frac{|\delta|}{2\lambda}}$). The nematic susceptibility in the symmetry broken phase can be obtained,
\begin{widetext}
\begin{align}
&\chi^M_{ij}(\Omega,{\bm p})=\frac{1}{C'(\Omega,\bm{p})}
\begin{pmatrix} 
-\frac{\kappa}{2F_2} p^2-|\delta| -\sin^2(2 \theta_p)\frac{1}{2\pi l_0^3} (\frac{i\Omega}{p}) &  i\Omega( \frac{2}{3\pi l_0^2} )-i \sin(2 \theta_p)\cos(2 \theta_p)\frac{1}{2\pi l_0^3} (\frac{i\Omega}{p})  \\ 
 -i\Omega (\frac{2}{3\pi l_0^2} )-i\sin(2 \theta_p)\cos(2 \theta_p)\frac{1}{2\pi l_0^3} (\frac{i\Omega}{p})& -\frac{\kappa}{2F_2}  p^2 -\cos^2(2 \theta_p)\frac{1}{2\pi l_0^3} (\frac{i\Omega}{p})
\end{pmatrix}
\end{align}
where
\begin{equation}
C'(\Omega,\bm{p})\approx (\frac{\kappa}{2F_2} p^2+ \frac{1}{2\pi l_0^3} \frac{i\Omega}{p}\sin^2(2 \theta_p))|\delta|,
\end{equation}
\end{widetext}

\subsection{Mode mixing in the Nematic phase and at quantum criticality}

In the theory of the nematic transition in a Fermi liquid of OFK, when approaching the criticality, $\delta \rightarrow 0$, the difference in the dynamics of the two polarizations becomes more noticeable. At criticality there is an underdamped longitudinal mode  $\omega_T\sim q^2$ and an overdamped transverse mode $\omega_L\sim iq^3$. In the nematic phase this mode becomes the (overdamped) Goldstone mode.  In our nematic criticality in the half-filled LL, due to the existence of the Berry phase term, the transverse and longitudinal modes are mixed, leading to the following modified dispersions for these collective modes
\begin{align}
\omega_1 &\sim \frac{\sqrt{\kappa V(0)} l_0}{\sqrt{2F_2 \pi }} q^2+ i \frac{l_0 V(0)}{2 }q^3,\nonumber\\
\omega_2 &\sim i \frac{ \kappa}{2 F_2 (\pi/l_0^2)^{3/2}} q^3
\end{align}
Thus, the transverse mode, $\omega_2$, remains overdamped (as in the OKF theory). In contrast, the longitudinal mode, $\omega_1$, is now underdamped (with $z=2$) only in the deep asymptotic long-wavelength  regime $q \to 0$, crossing over to an overdamped regime at larger values of $q$. This crossover can happen at long wavelengths if the range of the quadrupolar interaction is small.

\subsection{Electromagnetic Response and Spectral Peak}

In the nematic phase, the electromagnetic response (and the conductivity tensor) can be obtained after integrating out the gauge fluctuations. The calculation details of the electromagnetic response is presented in Appendix \ref{app:HLR}. Here we propose a possible experimental test  of the nematicity by measuring the spectrum of the conductivity $\sigma_{xx}$.
To this end, let us assume that we are in the deep nematic phase with a nonzero nematic order, say in the  $M_1$ component. The conductivity $\sigma_{xx}$ as a function of momentum is,
\begin{align}
\sigma_{xx}&=\frac{i}{\omega} \langle J_x J_x \rangle=-\omega \Pi_0(\bm{q},\omega) \frac{(1+M_1)}{4 D'(\bm{q},\omega)}, \nonumber\\
 D'(\bm{q},\omega)&= \Pi_0(\bm{q},\omega)\left(\Pi_2(\bm{q},\omega)-\frac{V(\bm {q})}{16\pi^2}\right)({\bm q}^2+M_1(q_x^2-q_y^2))\nonumber\\
&-\frac{1}{4}+\Pi^2_0(\bm{q},\omega)\omega^2
\end{align}
In the limit of $\omega/q\ll1$, the conductivity is approximately given by  
\begin{align}
\sigma_{xx}\sim \frac{\omega (1+M_1)/V(0)}{(1+M_1 \cos(2\theta_q))i\frac{\omega}{q V(0) l_0}+\frac{\omega^2}{V(0)^2 q^2}-q^2/4} 
\end{align}
Deep nematic phase where $M_1 \sim O(1)$, the damping term on the denominator will nearly vanish at $\theta_q=\pi/2$. Thus, by measuring  $\sigma_{xx}$ as a function of the angle $\theta_q$ of the direction of propagation measured form the nematic axis, one will find a resonance, i.e. a peak in the spectrum, at $\theta_q=\pi/2$.

\section{The Wen-Zee term in the CFL}
\label{sec:WZ}

From the Hall viscosity term, we expect that there may be the Wen-Zee type terms for the dynamical metric associated with the nematic order parameters.\cite{you2014,Maciejko2013} The Wen-Zee like term consists of two parts: a  parity-even term linear in nematic order parameter and in the gauge field, and a second term that is parity-odd, and is quadratic in nematic order parameter and linear in gauge field. At the mean field level of the CFL, which is  parity-even except for the Chern-Simons term, the composite fermions on the Fermi surface  generate only the parity-even part of the Wen-Zee term. However, upon the inclusion of the gauge field fluctuations, which are parity-odd, we will find the parity-odd part of the Wen-Zee term, as well as a modification of the parity-even part. 

We start by calculating the first part of the Wen-Zee term given by a Feynman  diagram that has one gauge leg and one nematic leg shown below in Fig. 3.
The coupling between gauge-invariant current, generated by the probe electromagnetic gauge fields $A_\mu$, and nematic order parameter $\bm M$  is given by
\begin{align}
&(-i)\frac{\delta \mathcal{S}}{\delta M_l \delta  A_k}(\bm{p},\Omega)\equiv \langle  M_l J_k\rangle ({\bm p},\Omega)\nonumber\\ 
=&i  \int_{\bm{q} ,\omega}~\textrm{tr}\Big[ \Pi_{ij}^{-1}(\omega,\bm{q}) \omega^2 \Pi_0 t_{l}  T^{-1}_{mi}(\omega+\Omega,\bm{q+p}) \omega^2 \Pi_0 Q_{k}\Big], \nonumber\\
=&\frac{1}{6\pi  \sqrt{3}}(\frac{\bar{\Lambda }m}{{\bar \rho}})^{1/3} \begin{pmatrix} 
 p_x\Omega & p_y\Omega\\
 -\Omega p_y  & p_x \Omega
\end{pmatrix}_{lk},
\end{align}
\begin{figure} [hbt] 
\begin{center}  
\includegraphics[width=0.4\textwidth]{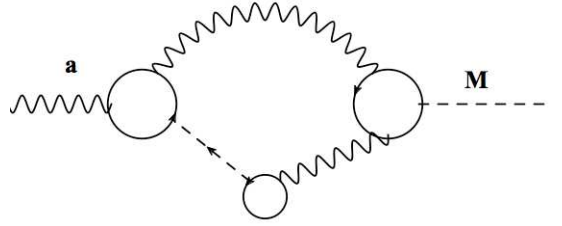}  
\caption{Leading parity-even  contribution to the Wen-Zee term. The wavy lines are gauge field propagators and the broken line is a nematic propagator. The blobs are composite fermion loops.}
\end{center}  
\label{fig:parity-even-mixing}
\end{figure}  
where, as before, $t_1=\sigma_3$ and $t_2=\sigma_1$, and where we used the notation $Q_x= I$ and $Q_y= -i \sigma_2$, and $T^{-1}_{mi}$ is given by
\begin{equation}
T^{-1}_{mi}=\langle a_m M_{i} \rangle=\frac{1}{\Pi_2 m q^2 \omega^2}
\begin{pmatrix} 
 q_x\omega & q_y\omega\\
 -\omega q_y  & q_x \omega
\end{pmatrix} 
\end{equation}
where we used the temporal gauge, $a_0=0$. This is the parity-even linear coupling between the nematic order parameter and the probe electromagnetic gauge field.

To calculate the parity-odd contributions to the Wen-Zee term, we calculate a Feynman diagram with one external probe gauge field leg and two nematic order parameter legs. Once again we choose the gauge $a_0=0$, and find the result
\begin{figure} [hbt] 
\begin{center}  
\includegraphics[width=0.35\textwidth]{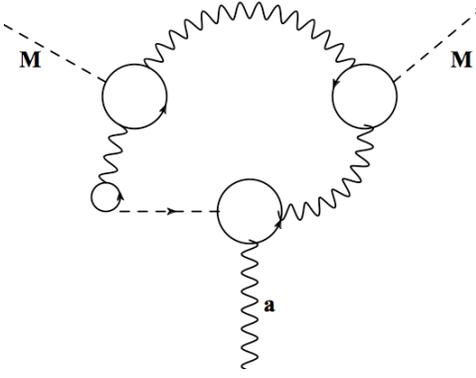}  
\caption{Parity-odd contribution to the Wen-Zee term at cubic level. The wavy lines are gauge field propagators and the broken line is a nematic propagator.}
\end{center}  
\label{fig:parity-odd-mixing}
\end{figure}
\begin{widetext}
\begin{align}
(-i)\frac{\delta \mathcal{S}}{\delta M_h \delta M_l \delta  A_k}&(\Omega_1,\Omega_2; \bm{p}_1,\bm{p}_2)\equiv \langle M_h M_l J_k\rangle (\Omega_1,\Omega_2; \bm{p}_1,\bm{p}_2)\nonumber\\
&= \int_{\bm{q},\omega}~\textrm{tr}\Big[ \Pi_{ij}^{-1}(\omega,\bm{q}) \omega^2 \Pi_0 \sigma_{h} \Pi_{ij}^{-1}(\omega+\Omega_1,\bm{q+p_1}) \omega^2 \Pi_0 \sigma_{l} 
T^{-1}_{mi}(\omega+\Omega_1+\Omega_2,\bm{q+p_1+p_2}) \omega^2 \Pi_0 Q_{k}\Big] \nonumber\\
&=\frac{1}{6  \pi \sqrt{3}}[(\frac{\bar{\Lambda}m}{{\bar \rho}})^{1/3} +\frac{\bar{\Lambda}m}{{\bar \rho}}]\epsilon^{hl} \epsilon^{\nu\mu k} p_1^{\nu}(p_1^{\mu}+p_2^{\mu})
\end{align}
\end{widetext}
which is manifestly  odd under parity and time reversal symmetries.

By combining the two contributions, we finally find a Wen-Zee term of the form
\begin{equation}
\mathcal{L}_{WZ}=\frac{1}{6\pi  \sqrt{3}}\Big[\left(\frac{\bar{\Lambda}m}{{\bar \rho}}\right)^{1/3} +\frac{\bar{\Lambda}m}{{\bar \rho}}\Big] \epsilon^{\mu \nu \rho}\omega_{\mu} \partial_{\nu} A_{\rho} \end{equation}
where $\bar \Lambda$ is the frequency UV cutoff of the CFL. Further reduce the expression by taking $\bar{\Lambda}m=l_0^{-2}$, we have,
\begin{equation}
\label{wz}
\mathcal{L}_{WZ}=\frac{2}{3\pi  \sqrt{3}} \epsilon^{\mu \nu \rho}\omega_{\mu} \partial_{\nu} A_{\rho} 
\end{equation}
Here we have denoted by $\omega_\mu$  the spin connection associated with the nematic fields,\cite{you2014}
\begin{align}
&\omega_{0}=\epsilon^{ij} M_i \partial_0 M_j \\\nonumber
&\omega_{x}=\epsilon^{ij} M_i \partial_x M_j-(\partial_x M_2-\partial_y M_1) \\\nonumber
&\omega_{y}= \epsilon^{ij} M_i \partial_y M_j +(\partial_x M_1+\partial_y M_2)
\end{align}
Similar to what we did for the calculation of the Berry phase term using a Ward Identity, the Wen-Zee term can also be derived from the Ward Identity, and the result is consistent with the diagrammatic calculation present here. The details of the calculation of Wen-Zee term from Ward Identity is presented in the Appendix \ref{app:details-WZ}. 

We should  stress that the connection between the Hall viscosity (given by the Berry phase term) and the Wen-Zee term in the CFL is not as straightforward as  in the incompressible FQH states. In the incompressible states, the universal coefficients $s$ of the Wen-Zee term is directly related to the Hall viscosity $\eta_H$, i.e., $\eta_H = \frac{s {\bar \rho}}{2}$.\cite{Read2009,Read2011,Cho-2014,Gromov-2015} However, in the compressible CFL state, the coefficients of the Wen-Zee term and the Hall viscosity do not have such relation because of the non-local nature of such responses in a compressible state. The coefficients appearing in the expressions are seemingly numeric constants, but it is important to remember that the coefficients are in fact the functions of the ratio $s = \frac{|\Omega|}{v_F |\bm{q}|}$ (not to be confused with the coefficient of the Wen-Zee term!), and are in general non-local in space and time. Instead, in the incompressible states where we can perform the well-defined gradient expansions due to the energy gap to all the excitations. Hence we find that in the CFL, the coefficients of the (seemingly local) Wen-Zee term and Hall viscosities are not universally related to each other.
\subsection{Nematic-Electric field correlator}
The Wen-Zee-type term Eq.\eqref{wz}, which couples the nematic field and the electromagnetic field, implies that the change in the nematic field will induce the electron quadrupole moment. Thus, one can measure the nematic susceptibility with respect to the external electric field $\chi_{M,E}=\frac{\partial T_i }{\partial E_j }$ as a function of momentum,
\begin{align}
&\chi_{M,E}=\frac{\delta^2 \mathcal{S}}{\delta M_i \delta E_j}(\bm{q})=\frac{\partial T_i }{\partial E_j },\nonumber\\
&=\frac{2}{3\pi  \sqrt{3}} \begin{pmatrix} 
 -q_x & q_y\\
 - q_y  & -q_x 
\end{pmatrix}. 
\end{align}
This susceptibility indicates that the energy-momentum current will be induced in the presence of the spatially-modulating electric field.

\section{General relation between the Hall viscosity and the hydrodynamic gauge theory in half-filled Landau levels}
\label{sec:hydro}

We have shown that, unlike the case of the incompressible FQH states, in the CFL the gauge field fluctuations contribute to the Hall viscosity of the half-filled Landau level.
In order to prove that the corrections to the Hall viscosity come from the fluctuation of the Chern Simons gauge fields, we start from CFL and dualize the CFL into a hydrodynamic gauge theory. 

If we couple our CFL to a smooth deformation of the background geometry (instead of to the nematic order parameter), we find that the composite fermions at the mean field level receive an orbital spin  from the flux attachment procedure.\cite{Cho-2014} Beyond this, if we include the gauge field fluctuation, the viscoelastic response of the CFL is related with the parity-odd electromagnetic response of the hydrodynamic gauge theory.\cite{Hoyos2012,Read2011} Since the original CFL is coupled to the fluctuating Chern-Simons gauge boson, the time-reversal odd electromagnetic response of the hydrodynamic gauge theory is always non-zero. Consequently, the CFL contains an additional contribution to the Hall viscosity arising from the fluctuating gauge field.

To see this, we start from the Chern-Simons theory of CFL where we attach two flux quanta of the Chern-Simons gauge field to the electrons to turn them into the composite fermions,
\begin{align}
&\mathcal{S}=\int d r^2 dt~ \Psi^{\dagger}\Big(\bm{r},t)(iD_t+\frac{1}{2m}{\bm D}^2+ \frac{g_{ij}}{2m} D_i D_j\Big)\Psi(\bm{r},t) \nonumber\\
&-\frac{1}{2}\int dr'^2 d r^2 dt~V(\bm{r-r'}) \rho ({r}) \rho({r'})\nonumber\\
&+\int d r^2 dt~ \frac{1}{8\pi} \epsilon^{\mu \nu \rho} a_{\mu}   \partial_{\nu}  a_{\rho} 
\end{align}
where $D_{\mu}=\partial_{\mu}+i(A_{\mu}+a_{\mu}+\omega_{\mu})$ is the covariant derivative.
Here, $\omega_\mu$ is the spin connection of the background metric $g_{ij}$, which is related to the local frame fields by  $g_{ij}=(e^a_j+\delta^a_j) (e^a_i+\delta^a_i)$. We only keep the leading orders in $e^a_i$ so that the distorted spatial metric is defined as,
\begin{align}
\delta g_{ij}=
\begin{pmatrix} 
  e_1^1     & e_1^2\\ 
  e_2^1 & e_2^2
\end{pmatrix}.
\end{align}
The composite fermion has  orbital spin $s=1$\cite{Cho-2014} so the covariant derivative of the composite fermion contains the spin connection with coefficient $1$ dictated by the orbital spin $s=1$. At the mean field level, the Chern-Simons flux cancels the external magnetic fields so we only need to consider the gauge fluctuation $\delta a$.

We can now perform the functional bosonization procedure, following Refs.[\onlinecite{Fradkin-1994,LeGuillou1997,Fradkin-1991,Chan-2013}], of our theory and introduce the hydrodynamic gauge field $b_\mu$,\cite{Wen-1995}
\begin{align}
&\mathcal{S}=\int d r^2 dt~ \Psi^{\dagger}(\bm{r},t)(iD_t+\frac{1}{2m}D_{i}D_{i}+  \frac{g_{ij}}{2m} D_i D_j)\Psi(\bm{r},t)\nonumber\\
& -\frac{1}{32\pi^2}\int d r'^2 d r^2 dt~ V(\bm{r-r'})  \delta b(\bm{r}) \delta b(\bm{r}') \nonumber\\
&-\int d r^2 dt~\frac{2}{4\pi} \epsilon^{\mu \nu \rho}  b_{\mu}   \partial_{\nu}  b_{\rho} 
+\int d r^2 dt~ \frac{1}{2\pi} \epsilon^{\mu \nu \rho} \delta a_{\mu}   \partial_{\nu}  b_{\rho}  
\end{align}
where $D_{\mu}=\partial_{\mu}+i(\delta a_{\mu}+\omega_{\mu})$ is the covariant derivative ,  and $\delta b(\bm{r}) = \partial_x \delta a_y(\bm{r})-\partial_y \delta a_x(\bm{r})$ is the fluctuating magnetic flux of the gauge field $\delta a_{\mu}$. 

Solving the saddle-point equation of $\delta a_\mu$, we find 
\begin{align}
\partial_{x} b_{0}-\partial_{0} b_{x}=&E^b_{x}=J^{CF}_{y},\nonumber\\
 \partial_{y} b_{0}-\partial_{0} b_{y}=&E^b_{y}=-J^{CF}_{x},\nonumber\\
\partial_x E^b_{x}=&T^{CF}_{12}=\Psi^{\dagger}  \frac{D_x D_y }{2m} \Psi,\nonumber\\
\partial_y E^b_{y}=&-T^{CF}_{21}=-\Psi^{\dagger}  \frac{D_y D_x}{2m}  \Psi
\end{align}

Now we turn on a momentum current of the composite fermion $T^{CF}_{12}$ in the system. Different from the parity-even Fermi surface where the metric only couples with the momentum current, the composite fermions carry intrinsic orbital spin so the spin connection appears in the covariant derivative. Hence,
\begin{align}
\frac{\delta \mathcal{S}}{\delta e_1^2}=T^{CF}_{12}+(\partial_{\mu} e^1_1) J^{CF}_{\mu}. 
\end{align}
The composite fermion current $T^{CF}_{12}$ is bound with the spatial derivative of electric field of $b_\mu$ as $\partial_x E^b_{x}$. Once we have a nonzero $\partial_x E^b_{x}$, a polarized charge density appears. If the hydrodynamic gauge theory of the gauge field $b_\mu$ contains a term like $ \beta (\nabla \cdot E^b)B^b$,  the polarized charge density associated with the field  $b_\mu$  acts as a magnetic moment which couples with magnetic flux of $b_\mu$. Thus, we have
\begin{align}
\frac{\delta \mathcal{S}}{\delta (\partial_x  b_y) }=\beta \partial_x (\partial_{x} b_{0}-\partial_{0} b_{x})
\end{align}
By solving the equation of motion for the hydrodynamic gauge field,
\begin{align}
&\frac{\delta \mathcal{S}}{\delta (\partial_x  b_y) }=\partial_0 \frac{\delta \mathcal{S}}{\delta \partial_0(\partial_x  b_y) }\nonumber\\
&\frac{\delta \mathcal{S}}{\delta \partial_0(\partial_x  b_y) }=\frac{\delta \mathcal{S}}{\delta \partial_x(\partial_0  b_y) }=\frac{\delta \mathcal{S}}{\delta T_{11} }=e_1^{1}
\end{align}
We finally have
\begin{align}
&\frac{\delta \mathcal{S}}{\delta (\partial_x  b_y) }=\partial_0 e_1^{1}=\beta \partial_x (\partial_{x} b_{0}-\partial_{0} b_{x}). 
\end{align}
Thus the Hall viscosity can be read off from 
\begin{align}
\frac{\delta \mathcal{S}}{\delta e_1^2}=\frac{1}{\beta}\partial_0 e_1^{1}+(\partial_{\mu} e^1_1) J^{CF}_{\mu}=(\rho+\frac{1}{\beta}) \partial_0 e_1^{1}, 
\end{align}
i.e., the Hall viscosity is $(\rho+\frac{1}{\beta})$ in which $\rho$ is due to the flux attachment,\cite{Cho-2014} and $1/\beta$ is due to the parity-odd fluctuations of the hydrodynamic gauge field. The parity-odd viscosity has two contributions:  one is the intrinsic orbital spin that the composite fermion carries, and the other is through the term $ (\nabla \cdot E^b)B^b$ of the hydrodynamic gauge field. Hence, if the dual hydrodynamic gauge theory has the parity-odd  $\beta (\nabla \cdot E^b)B^b$ term, we expect that there should be additional Hall viscosity from the gauge fluctuation. The theory of the $b_\mu$ field can be obtained by integrating out the composite fermion surface and gauge fluctuation of the gapless $a_\mu$. However, if the effective theory of $a_\mu$ is gapless and nonlocal, the coefficient $\beta$ for the term $(\nabla \cdot E^b)B^b$ will be non-universal (and depend on the UV cutoff in a singular way). This is in accordance with our result for the CFL.

\section{Connection to Experiments}
\label{Experiment}

In this section, we discuss the connections between our results with the experiments\cite{lilly-1999,du-1999,Sambandamurthy-2008,Samkharadze-2015} on  half-filled Landau levels with $N>1$. These experiments show that in half-filled Landau levels with $N\geq 2$, there is a spectacular  anisotropy in the longitudinal transport with ratios of the resistances as large as $R_{xx}/R_{yy} \sim 3,500$ at the lowest temperatures.  The anisotropy in the longitudinal resistivities, expressed in the difference $\rho_{xx}- \rho_{yy}$,  raises very rapidly below a critical temperature $T_c \sim 65$mK (for $N=2$). These results were  originally interpreted as the signature of a striped phase,\cite{moessner-1996,koulakov-1996,koulakov-1996b} an unidirectional charge-density-wave state which breaks translation symmetry (and, necessarily, rotational symmetry.)  However,  further transport experiments showed that the $I-V$ (current-voltage) curves were metallic and showed a linear behavior at low bias voltages.\cite{lilly-1999} In contrast, a charge-density-wave (CDW) would have exhibited non-linear $I-V$ curves with a sharp onset at a critical voltage. Extremely sharp onset behavior has been seen indeed in the reentrant integer quantum Hall regime away from the center of the Landau level and has been interpreted as evidence for a ``bubble phase" (i.e. a bidirectional CDW.) Moreover, in the reentrant IQH regime the experiments show
broadband noise in the current, which are observed in many CDW phases, but which is absent  the anisotropic half-filled Landau levels. 

For these reasons, the experiments in the center of the Landau levels (with $N\geq 2$) have been interpreted instead as evidence for a {\em nematic} phase phase, i.e. a uniform and compressible phase of the 2DEG with a strong anisotropy (for a review see Ref. [\onlinecite{Fradkin-2010}].)
This interpretation is further supported by comparing the anisotropy in transport data with a simple model of a nematic, a two-dimensional classical $XY$ model for  a director order parameter. By menas of Monte Carlo calculations it was found that indeed this model fits really well the transport anisotropy data.\cite{Fradkin-2000, cooper-2001}  Furthermore, these fits show that there is a very low energy scale for the native anisotropy is of the order of $3-5$mK, which is presumably related to the coupling of the 2DEG to the underlying lattice. Although the experiments cannot exclude the possibility that the ground state at $T=0$ is actually a possible striped state, which will be melted thermally into a nematic state at finite temperature, the absence of any evidence of stripiness at the lowest temperatures seems to imply that the ground state is a nematic (which, most likely, should be regarded as a quantum melted striped phase.)
A detailed review on these experiments (prior to 2010) and their interpretation can be found in Ref. [\onlinecite{Fradkin-2010}].

In this section we   focus on a relatively recent set of experiments on  radio-frequency conductivity measurements in the $N=2$ Landau level near filling fraction $\nu \approx 9/2$ by Sambandamurthy \textit{et.al.}\cite{Sambandamurthy-2008} which, from our perspective, can be naturally interpreted as the following. In this experiment,\cite{Sambandamurthy-2008} they observe an anisotropy in the longitudinal conductivities as well as a resonant peak of the radio-frequency longitudinal conductivity along the hard direction of transport, say $\sigma_{xx}$, with a frequency of $100$ MHz. This $100$ MHz resonance was originally interpreted as evidence for the existence of a  pinning mode of a stripe state.\cite{Cote-2000}  Given that the $I-V$ curves are linear, and hence that there is no evidence of translation symmetry breaking, it is natural to seek a nematic explanation for this energy gap. In a nematic state in the continuum there would not be a gap. On the other hand, the transport anisotropy experiments show that the anisotropy saturates below $T \sim 20$mK. Tilted-field experiments\cite{cooper-2001,Pollanen-2015} and the fits to the $XY$ model with a weak symmetry-breaking field,\cite{Fradkin-2000} show that the energy scale is of the order of $3-5$mK, which is quite comparable with a resonant frequency of $100$nHz as seen by Sambandamurthy \textit{et.al.}\cite{Sambandamurthy-2008} Thus, we are led to the interpretation that the radio-frequency conductivity measurements are detecting a nematic Goldstone mode gapped out by the coupling to the lattice which induces a term that breaks the continuous rotational symmetry to the $C_4$ symmetry of the lattice (i.e. of the surface on which the 2DEG is confined.) It is also important to note that the resonant peak seen in by Sambandamurthy \textit{et.al.} behaves remarkably close to what is seen in the transport anisotropy in the DC experiments.

Finally, we should note that transport experiments in the $N=1$ Landau level have shown a close connection between the nematic state and the paired FQH state at $\nu=5/2$. Indeed, earlier experiments\cite{Xia2010} showed that by tilting the magnetic field the FQH state at $\nu=5/2$ is destroyed and that the resulting compressible state behaves much like an anisotropic HLR state. A relatively recent experiment\cite{Xia2011} has also given evidence of a nematic state in the $\nu=7/3$ FQH plateau (presumably a Laughlin-type state) and was interpreted as such in several theory papers.\cite{Mulligan2010,Mulligan2011,Maciejko2013,you2014} Very recent transport experiments
by Samkharadze \textit{et.al.},\cite{Samkharadze-2015} found that, by applying a sufficiently large external hydrostatic pressure, the \textit{incompressible} isotropic FQH state at $\nu = \frac{5}{2}$ spontaneously gives a way to the compressible anisotropic phase. This is possible since the pressure tunes the width of the quantum well and thus tunes the effective interaction between the electrons, as the form and spread of the electron wavefunction will be varied as the width is changed. At the critical pressure, it is found that the rotational symmetry is spontaneously broken and the anisotropy in conductivities develops. Though this transition is between \textit{incompressible} QH state and \textit{compressible} nematic CFL instead of the transition from the compressible isotropic state, namely CFL, to the compressible anisotropic state, this result gives a strong hint  that tuning an external parameter, such as pressure,  the effective electron interaction such as $F_2$ in this paper can be tuned  to find the transition that we have studied in this paper.  

\section{conclusions and outlook}

\label{sec:conclusions}

In this work, we considered the problem of a 2DEG in a  half-filled Landau level with strong quadrupolar interactions. We  mapped the fermion theory into a composite Fermi liquid coupled to a gauge field with a Chern Simons term, extended by a quadrupolar interaction. Both the nematic fluctuations and gauge fluctuations  are found to soften the Fermi surface and drive the system into a non-Fermi liquid state. We  started from the nematic Fermi liquid theory corrected by the fluctuations of the Chern-Simons gauge boson fluctuations and looked at the nematic instability of the non-Fermi liquid. The nematic fluctuations are Landau-damped by the non-Fermi liquid state and thus the dynamical critical exponent  is $z=3$.  

The nematic theory was also shown to contain a Berry phase term arising from the  gauge field fluctuations, which  correct the nematic correlator. This non-zero Berry phase term suggests the gauge fluctuation would also contribute to the Hall viscosity of the half-filled Landau level. The resulting  odd viscosity in this gapless system is non-universal. 
In addition, inside the nematic phase, the nematic vortex current couples with the gauge field. This also demonstrates that the half-filled Landau level orbital spin is not exactly equal to $s=1$ as both gauge fluctuation and orbital spin of the CFL in the mean field level have separate contribution to Wen-Zee term. The computation of the Hall viscosity was confirmed by an argument based on a set of Ward identities. 

The theory of the nematic composite Fermi liquid presented here is based on the concept of flux attachment, i.e. on the equivalency between different theories of interacting fermions in two space dimensions to a theory of composite fermions coupled to a gauge field. This approach has been known for a long time to give the correct universal properties of the FQH states,\cite{Lopez-1991} including subtle responses to changes in the external geometry.\cite{Cho-2014,Gromov-2015} These theories are also known to give a qualitative description of the compressible phases, i.e. the HLR theory.\cite{Halperin-1993} However, it is also well known that the mean field approximations based on this mapping involve a large amount of Landau level mixing (even in the limit of a very large magnetic field) which becomes extreme in the compressible states. A symptom of these problems is the lack of particle-hole symmetry even in the limit in which all excited Landau levels are projected out. These difficulties  have been the focus of intense recent work,\cite{Son-2015,Barkeshli-2015,Wang-2015,Geraedts-2015,Metlitski-2015} already noted in the Introduction. This approach proposes to describe the half-filled Landau level, instead, as proximate to a theory of Dirac fermions ``dually'' coupled to a dynamical gauge field. The problem of the possible connection between the ``more conventional'' Chern-Simons approach and these recent proposals is at present unclear, including how they may relate to the nematic and paired states. We will consider these connections in a separate publication.

Finally we note that there have not been systematic numerical studies of the nematic transition in 2DEGs in large magnetic fields. Most of the existing studies of rotational (and translational) symmetry breaking have been done either by means of  finite-size diagonalizations\cite{Haldane-2000} (done on small systems with toroidal boundary conditions which break rotational invariance explicitly) or with projected wave functions \cite{Lee-2015} or with variational wave functions ,\cite{doan-2007} mostly done on relatively small systems on the sphere (which also poses problems for an order that breaks rotational invariance). A more careful numerical study of this problem is clearly needed.

\begin{acknowledgements}
This work was supported in part by the National Science Foundation through grants  DMR-1408713 (Y.Y,E.F) at the University of Illinois, and PHY11-25915 at the Kavli Institute for Theoretical Physics  (KITP) (Y.Y.), and the Brain Korea 21 PLUS Project of Korea Government (G.Y.C). Y.Y thanks the KITP Graduate fellowship program for support and G.Y.C thanks ICMT for partial support and hospitality.
\end{acknowledgements}


\appendix
\section{The isotropic-Nematic quantum phase transition in a Fermi Liquid}
\label{app:OKF}

The problem of the isotropic-nematic quantum phase transition by a Pomeranchuk instability in a Fermi liquid (without a background lattice) was  studied by Oganesyan, Kivelson and Fradkin\cite{Oganesyan-2001} whose work we follow in detail. The lattice version of this problem was studied by several authors.\cite{halboth-2000,metzner-2003,khavkine-2004} For a review of electronic nematic phases see Ref.[\onlinecite{Fradkin-2010}]. Here we will use a  perturbative approach, following the standard work of Hertz\cite{Hertz-1976} and Millis\cite{Millis-1993} (for a review see Ref.[\onlinecite{Sachdev-1999}]). The full non-perturbative behavior of  Fermi fluid is not fully understood and has been the focus of considerable work, both analytic\cite{Metlitski-2010} and, more recently, numerical.\cite{Schattner-2015} The problem of the quantum phase transition to an electron nematic state from a charge stripe state has not been studied as much (see, however, Ref.[\onlinecite{sun-2008}].)

We start from the isotropic FL described by the free-fermion action $\mathcal{S}_0$ in two space dimensions for spinless fermions (spin will play no role here),
\begin{align}
\mathcal{S}_0 = \int  d^2 r dt \Psi^{\dagger}(\bm{r},t) \Big[ i\partial_t  + \frac{\bm{\nabla}^2}{2m} + \mu  \Big]\Psi(\bm{r}, t),
\label{FL1}
\end{align}
in which $\mu$ is the chemical potential and $\Psi$ is the spinless fermionic field.
This theory $\mathcal{S}_0$ is invariant under arbitrary spatial rotations. It is known that this Fermi liquid is stable to all \textit{infinitesimal} interaction except the superconducting (BCS) channel.\cite{Shankar-1994,Polchinski-1993} Hence, by excluding the pairing-channel, the only way to introduce any qualitative and quantitative change is to turn on the interaction beyond some  finite strength. Hereafter, we will ignore the pairing instability and concentrate on phase transitions only in the particle-hole channel (although the nematic quantum criticality can lead to a superconducting state.\cite{Metlitski-2015b,Raghu-2015}) 
In addition, Oganesyan and coworkers,\cite{Oganesyan-2001}  found that in order to stabilize the nematic ground state it is necessary to include in the free fermion Hamiltonian terms in the dispersion relation that are at least cubic in the momentum relative to the Fermi momentum. Such terms  are explicitly irrelevant in the Landau FL phase. Although in this section we will not include these terms explicitly, we will make them explicit in the theory of the nematic CFL of Section \ref{sec:Nematic-CFL}. 

We are interested in the process of the spontaneous breaking of the rotational symmetry in a FL. The most obvious way to break the rotational symmetry is the spontaneous distortion of the Fermi surface. In this paper, we are mainly interested in the distortion in the $d$-wave channel, i.e., the quadrupolar channel. The spontaneous symmetry breaking transition is thus induced naturally by turning on the  strength of the Landau parameter with an attractive coupling $F_2$ for the quadrupolar interaction, represented by a term $\mathcal{S}_Q$ in the full action of the form\cite{Oganesyan-2001}
\begin{align}
\mathcal{S}_Q = \frac{1}{2} \int dt d^2r d^2r' \, F_2(|\bm{r}-\bm{r}'|) \textrm{tr}\Big( \hat{Q}(\bm{r},t)\cdot \hat{Q}(\bm{r}',t) \Big),
\label{FL2}
\end{align}
where $ F_2(|\bm{r}-\bm{r}'|)$ is  a short-ranged quadrupolar interaction,
\begin{equation}
F_2(|\bm{r}-\bm{r}'|)=\int \frac{d^2 {q}}{(2\pi)^2} \frac{F_2}{1+ \kappa \bm{q}^2}
\label{eq:F2}
\end{equation}
where $\kappa^{-1}$ is the range of the quadrupolar interaction and $F_2$ is the quadrupolar coupling.
In Eq.\eqref{FL2}  denoted by $\hat{Q}(\bm{r})$  the electronic quadrupolar density defined in the OFK paper,
\begin{align}
\hat{Q}(\bm{r}) = \frac{1}{ k_F^2}\Psi^{\dagger}(\bm{r})
\begin{pmatrix} 
\partial_x^2 - \partial_y^2 & 2 \partial_x \partial_y \\
2\partial_x \partial_y & \partial_y^2 -\partial_x^2  
\end{pmatrix}
\Psi(\bm{r}). 
\label{eq:SQ}
\end{align}
Note that the electronic quadrupolar density in the OFK paper is different from our definition in Section \ref{sec:Nematic-CFL} up to $k_F^2$.

It is clear by dimensional counting  that the interaction Eq.\eqref{FL2} is irrelevant at the Fermi liquid fixed point of Eq.\eqref{FL1}, and thus we will need the \textit{finite} strength of $F_2$ in Eq.\eqref{FL2} to be large enough (and attractive) to drive a phase transition out of the isotropic Fermi liquid Eq.\eqref{FL1}.

To understand the quantum phase transition better, we decouple the interaction term of Eq.\eqref{eq:SQ} by means of a  Hubbard-Stratonovich transformation,  and replace the quartic form of the action $\mathcal{S}_Q$ by another one in which the nematic order parameters are coupled linearly to two real Hubbard-Stratonovich fields, $M_1$ and $M_2$. After this is done, Eq.\eqref{eq:SQ} becomes
\begin{align}
\mathcal{L}_Q &=   \frac{M_{1}(\bm{r},t)}{k_F^2} \Psi^{\dagger}(\bm{r},t) (\partial_x^2 - \partial_y^2)\Psi(\bm{r},t) \nonumber\\
&+ \frac{M_{2}(\bm{r},t)}{k_F^2} \Psi^{\dagger}(\bm{r},t)  (2\partial_x\partial_y) \Psi(\bm{r},t)  \nonumber\\ 
& - \frac{1}{2 F_2} \left[|\bm{M}(\bm{r},t)|^2 + \kappa  \left(\bm{\nabla} \bm{M}(\bm{r},t) \right)^2\right]
\label{eq:HS}
\end{align}
where we introduced the {\em director} field $\bm{M}=(M_1,M_2)$.

Hence we obtain a theory of the fermion nematic order parameter coupled to the Hubbard-Stratonovich fields $M_1$ and $M_2$. In  momentum and frequency space the  action becomes
\begin{widetext}
\begin{align}
\mathcal{S}  &=  \int_{\bm{k},\omega} \Big[\Psi^{\dagger}(\bm{k},\omega) \Big(\omega - \frac{\bm{k}^2 - k_F^2}{2m} \Big)\Psi(\bm{k},\omega)
- \frac{1}{2 F_2} \left(1+\kappa \bm{k}^2 \Big) |\bm{M}(\bm{k},\omega)|^2  \right) \Big] \nonumber\\
&+\int_{\bm{k},\omega}  \int_{\bm{q},\Omega} \Psi^\dagger(\bm{k}+\bm{q},\omega+\Omega) 
\Big[\frac{M_1(\bm{q},\Omega)}{k_F^2} (k_x^2-k_y^2) + \frac{M_2(\bm{q},\Omega)}{k_F^2} (2k_x k_y) \Big] \Psi(\bm{k},\omega), 
\end{align}
\end{widetext}
Here we used the the short-hand notation $\int_{\bm{q},\omega} = \int \frac{d\omega d^2 q}{(2\pi)^3}$ and have set the Fermi momentum to be $k_F = \sqrt{2m \mu}$, and the chemical potential $\mu$ is the Fermi energy. 

After the Hubbard-Stratonovich transformation, we proceed to integrate out the fermions  to obtain the effective action for the order parameters $M_{1}$ and  $M_{2}$. Close to the quantum phase transition to the nematic state we can approximate the effective action by a Landau  expansion in powers  of the nematic order parameter fields. To quadratic and quartic order one finds
\begin{align}
S_{n} = \frac{1}{2 N_F}& \int d\omega d^2q ~ M_{i}(\bm{q},\omega) \mathcal{L}_{ij}(\bm{q}, \omega) M_{j} (\bm{q}, \omega) \nonumber\\
-& \int d^2r dt \left[ \frac{\kappa}{2 F_2}  \left(\bm{\nabla} \bm{M} \right)^2+ \lambda \bm{M}^4 \right]
\label{eq:OKF}
\end{align}
where\cite{Oganesyan-2001} $\lambda=(3\alpha N_F |F_2|^3)/(8E_F^2)$, and $\alpha$ is the coefficient of the quartic term in the single-particle dispersion.
The inverse of the propagator of the order parameters, $\mathcal{L}_{ij}(\bm{q}, \omega)$, contains the information about quantum critical dynamics of the order parameters. The analytic form of $\mathcal{L}_{ij}(\bm{q}, \omega)$ can be  obtained up to  one-loop correction in the fermions. The result is\cite{Oganesyan-2001}
\begin{equation}
\mathcal{L}_{ij}(\bm{q}, \omega)=\delta_{ij}(\frac{\kappa}{2 F_2} \bm{q}^2+\delta)+\mathcal{M}_{ij}(s,\phi),
\label{eq:Seff-one-loop}
\end{equation}
where $\delta=-\frac{1}{2}-\frac{1}{N_F F_2}$ parametrizes the distance  from the nematic quantum critical point at $F_{2}^{*} = -\frac{N_F}{2}$ (the Pomeranchuk transition), $s=\frac{\omega}{v_F q}$, $v_F = \frac{k_F}{m}$ is the Fermi velocity, and $\phi$ is the polar angle of the momentum $\bm{q}$.  
The matrix kernel $\mathcal{M}_{ij}(s,\phi)$ is given by
\begin{equation}
\mathcal{M}_{ij}(s,\phi)=\frac{s}{2}
\begin{pmatrix} 
B(s)+A(s)\cos(4\phi)   & A(s)\sin(4\phi) \\ 
  A(s)\sin(4\phi) & B(s)-A(s)\cos(4\phi) 
\end{pmatrix},
\label{eq:calM}
\end{equation}
where
\begin{equation}
B(s)=\frac{1}{\sqrt{s^2-1}},\qquad A(s)=B(s)(\sqrt{s^2-1}-s)^4,
\end{equation}

Since we are interested in the dynamics of the asymptotic regime $s = \frac{\omega}{v_F q} \ll 1 $, we further expand the functions  $A(s)$ and $B(s)$ for small $s$ around $s=0$ and take $\phi=0$ to find   
\begin{align}
&\mathcal{L}_{ij}=\begin{pmatrix} 
-i\frac{\omega}{v_F q}+\frac{\kappa}{2 F_2} q^2 +\delta  & 0 \\ 
  0 & -(\frac{\omega}{v_F q})^2+\frac{\kappa}{2 F_2} q^2 +\delta
\end{pmatrix}, 
\end{align}
where $q=|\bm{q}|$. This result of Oganesyan {\it et al.}\cite{Oganesyan-2001}  shows that the quantum dynamical exponent is $z=3$. The finite density of states at the Fermi surface is the origin of this strong Landau damping of this longitudinal critical mode. 

Provided the rotational symmetry of the system is not explicitly broken, inside the nematic phase there is a Goldstone mode associated with the spontaneously broken rotational invariance. In this metallic system, the Goldstone mode is Landau damped. Oganesyan {\it et al.} showed that this overdamped Goldstone mode leads, to lowest order in perturbation theory,  to a quasiparticle self-energy whose imaginary part scales as $\Sigma''(\omega) \propto |\omega|^{2/3}$ and, consequently, to the breakdown of the quasiparticle picture and to non-Fermi liquid behavior. However, in the case of the 2DEG, the continuous rotations symmetry is broken down to the $C_4$ point group symmetry of the surface. Although this explicit symmetry-breaking is very weak, it results in a finite (but small) energy gap for the nematic Goldstone mode. The results summarized above hole above this energy scale.

It is important to emphasize  that in this picture  the quadrupolar interaction of Eq.\eqref{FL2} drives the quantum phase transition to the nematic state, if the coupling constant $F_2$ exceeds the critical value. We will see in Section \ref{sec:Nematic-CFL} that the same interaction will induce the nematic quantum phase transition in the CFL. 

\section{The HLR Composite Fermi Liquid}
\label{app:HLR}

Now we review briefly the physics of the composite Fermi liquid (CFL)  of Halperin, Lee and Read,\cite{Read-1994,Halperin-1993,kim1994instantons,Rezayi-1994} which is another key ingredient of our analysis. Before flux attachment, the action for the HLR compressible CFL is
\begin{align}
\mathcal{S} &= \int dt d^2r  \Psi^{\dagger}(\bm{r},t) \Big[ iD_t  + \frac{\bm{D}^2}{2m} + \mu  \Big]\Psi(\bm{r}, t),\nonumber\\ 
& -\frac{1}{2} \int dt d^2r d^2r' ~V(|\bm{r}-\bm{r}'|) \delta \rho(\bm{r}, t)\delta \rho(\bm{r}', t)
\end{align}
in which we have introduced the covariant derivative $D_{\mu} = \partial_\mu + iA_{\mu}$, with $\mu=t,x,y$, where  $A_\mu$ is the external electromagnetic gauge field. Here $V(|\bm{r}-\bm{r}'|)$ represents the density-density interaction between the electrons, i.e., $\delta \rho(\bm{r}) = \Psi^{\dagger}(\bm{r})\Psi(\bm{r})- \bar{\rho}$ is the local deviation of the electronic density  from the average  $\bar{\rho}$.  In Section \ref{sec:Nematic-CFL} we  will also include a quadrupolar interaction discussed in Section \ref{app:OKF}.  

For a half-filled Landau level,  the average density $\bar{\rho}$ of the electrons and the uniform magnetic field $B$, are related by the filling factor $\nu =2\pi  \frac{\bar{\rho}}{B} = \frac{1}{2}$. The HLR theory also applies to the other compressible states at filling factors $\nu=1/2n$. In this paper we will focus on the case $\nu=1/2$. We will also include a probe (and unquantized)  component of the electromagnetic gauge field, which we denote by  $\delta A_\mu$. The total electromagnetic field is  $A_\mu = \bar{A}_\mu + \delta A_\mu$. 

Now we perform the flux attachment transformation 
suitable for $\nu = \frac{1}{2}$, i.e. we will attach two flux quanta to each fermion.
The flux attachment is implemented by coupling the fermions to a statistical gauge field $a_\mu$ whose action is a  Chern-Simons term. The total action of the transformed, composite, fermion is\cite{Lopez-1991,Halperin-1993}
\begin{align}
\mathcal{S} &= \int dt d^2 r  ~ \Psi^{\dagger}(\bm{r},t) \Big[ iD_t  + \frac{\bm{D}^2}{2m} + \mu  \Big]\Psi(\bm{r}, t) \nonumber\\ 
& -\frac{1}{2} \int dt d^2 r d^2 r' ~V(|\bm{r}-\bm{r}'|) \delta \rho(\bm{r}, t)\delta \rho(\bm{r}', t), \nonumber\\ 
& + \int dt d^2 r \frac{1}{8\pi} \varepsilon^{\mu\nu\lambda} a_{\mu}\partial_\nu a_\lambda,
\end{align}
Here $D_{\mu} = \partial_\mu + iA_{\mu}+ia_\mu$ (where $\mu=t,x,y$), is the covariant derivative required by a gauge-invariant (minimal) coupling of the fermions to  the electromagnetic gauge field $A_\mu$ and to the Chern-Simons gauge field $a_\mu$. 

We now perform the average field approximation in which the average part $\bar{A}_\mu$ of the electromagnetic gauge field is cancelled by the average part $\bar{a}_\mu$ of the Chern-Simons gauge field, $\bar{A}_\mu + \bar{a}_\mu = 0$. 
After this approximation, we end up with the effective theory 
\begin{align}
\mathcal{S} &= \int dt d^2 r  \Psi^{\dagger}(\bm{r},t) \Big[ iD_t  + \frac{\bm{D}^2}{2m} + \mu  \Big]\Psi(\bm{r}, t)
\nonumber\\
&+ \int dt d^2 r  \frac{1}{8\pi} \varepsilon^{\mu\nu\lambda} \delta a_{\mu}\partial_\nu \delta a_\lambda,\nonumber\\ 
& -\frac{1}{2} \int dt d^2 r d^2 r' ~V(|\bm{r}-\bm{r}'|) \delta \rho(\bm{r}, t)\delta \rho(\bm{r}', t), 
\end{align}
with $D_{\mu} = \partial_\mu + i\delta A_{\mu}+i\delta a_\mu$, where we set $a_\mu = \bar{a}_\mu + \delta a_\mu$. 

The Chern-Simons gauge theory is a topological field theory.\cite{Witten-1989} At the local level, its content is a a set of commutation relations between the spatial components of the gauge field, and a constraint on the space of states (a Gauss law) which in this case reduces to a constraint between the charge density, $\delta \rho(\bm{r},t)$ and the local flux of the Chern-Simons gauge fields, 
\begin{align}
\delta \rho(\bm{r},t) = \frac{1}{4\pi} \varepsilon^{tij}\partial_i \delta a_j(\bm{r},t) = \frac{1}{4\pi} \delta b(\bm{r},t),
\label{eq:Gauss}
\end{align} 
as an operator identity in the Hilbert space of gauge-invariant states.

Finally, upon a shift of the Chern-Simons gauge field, $\delta a_\mu \to \delta a_\mu - \delta A_\mu$, we can write the action in the form 
\begin{align}
\mathcal{S} &= \int dt d^2 r  \Psi^{\dagger}(\bm{r},t) \Big[ iD_t  +\frac{\bm{D}^2}{2m} + \mu  \Big]\Psi(\bm{r}, t),\nonumber\\ 
& -\frac{1}{2} \int dt d^2 r d^2 r' ~\frac{1}{16\pi^2} V(|\bm{r}-\bm{r}'|) \delta b(\bm{r}, t)\delta b(\bm{r}', t), \nonumber\\ 
& +  \int dt d^2 r \frac{1}{8\pi}\varepsilon^{\mu\nu\lambda} (\delta a_{\mu}-\delta A_\mu)\partial_\nu (\delta a_\lambda-\delta A_\lambda), 
\label{eq:CFL-action}
\end{align}
where now the covariant derivative is $D_{\mu} = \partial_\mu + i\delta a_\mu$. Here we have used the Gauss law constraint of Eq.\eqref{eq:Gauss}, to write the density-density  interaction in terms of the fluctuations of the gauge {\em flux} $\delta b(\bm{r},t)$, resulting in a flux-flux coupling for the gauge fields.\cite{Lopez-1991,Lopez-1993} We will see in Section \ref{sec:Nematic-CFL} that for the quadrupolar interaction this identity does not apply and the form of the interaction is more complex.

Therefore, the flux attachment transformation maps a half-filled Landau level to a system of composite fermions at finite density coupled to a dynamical gauge field (with vanishing average) with a Chern-Simons term (and a Maxwell-like term as well).  The Fermi momentum of the composite fermions is $k_F=(4\pi {\bar \rho})^{1/2}=B^{1/2}=\ell_0^{-1}$, where $\ell_0$ is the magnetic length. Since the composite fermions do not experience the magnetic field (on average), the composite fermions  form a FL, provided that the coupling to the gauge field $\delta a_\mu$ can be neglected. This is a gapless state, the HLR composite Fermi liquid.\cite{Halperin-1993} The crucial question is what tis he fate of the FL  if the coupling is included. In fact, it turns out that the coupling of the composite fermions to the fluctuating gauge field destroys completely the well-defined composite fermion quasiparticles on the Fermi surface and leads to a non-Fermi liquid state .

Since the CFL action of Eq.\eqref{eq:CFL-action} is quadratic in the composite fermion fields, we can proceed to integrate out the composite fermions,  and obtain an effective action for the fluctuating gauge fields $\delta a_\mu$. HLR showed that, at the one loop (RPA) level, the fluctuation $\delta a_\mu$ of the gauge field  experiences a strong Landau damping due to the finite density of states of electron-like excitations at the Fermi surface. The damping manifests in the polarization bubble of the gauge field when the fermion is integrated out at the one-loop level (with the free fermion propagator). To quadratic order in the gauge fields, their effective action in the CFL state is,
\begin{align}
\mathcal{S}_{\textrm{CFL}}[\delta a_\mu,& \delta A_\mu]=\nonumber\\
-\frac{1}{2}& \int_{\bm{q},\omega}  \delta a_{\mu}(\bm{q},\omega) \Pi_{\mu\nu}(\bm{q},\omega) \delta a_{\nu}(-\bm{q},-\omega)\nonumber\\
+\frac{1}{8\pi} &\int dt  d^2 r \varepsilon^{\mu\nu\lambda} (\delta a_{\mu}-\delta A_\mu)\partial_\nu (\delta a_\lambda-\delta A_\lambda), 
\label{effectiveCFL}
\end{align}
where $ \Pi_{\mu\nu}(\bm{q},\omega)$ is the polarization tensor of the CFL, 
\begin{widetext}
\begin{equation}
\Pi_{\mu\nu}(\bm{q},\omega) = \frac{1}{2\pi}
\begin{pmatrix} 
q^2 \Pi_0(\bm{q},\omega) & q_x\omega \Pi_0(\bm{q},\omega)  & q_y\omega \Pi_0(\bm{q},\omega) \\
q_x\omega \Pi_0(\bm{q},\omega)   & \omega^2 \Pi_0(\bm{q},\omega)-q_y^2\Big(\Pi_2(\bm{q},\omega)+V(q)\Big)  & q_yq_x\Big(\Pi_2(\bm{q},\omega)+V(q)\Big) \\
q_y\omega \Pi_0(\bm{q},\omega)   & q_yq_x\Big( \Pi_2(\bm{q},\omega)+V(q)\Big)  &\omega^2 \Pi_0(\bm{q},\omega)-q_x^2\Big( \Pi_2(\bm{q},\omega)+V(q)\Big)
\end{pmatrix},
\label{eq:Pimunu}
\end{equation}
\end{widetext}
where $V(q)$ is the Fourier transform of the interaction. The functions $\Pi_0(\bm{q},\omega)$ and $\Pi_2(\bm{q},\omega)$ are
\begin{equation}
\Pi_0(\bm{q},\omega) =\frac{m}{q^2}-\frac{i|\omega| }{ \sqrt{\bar \rho} |q|^3},\quad
\Pi_2(\bm{q},\omega) =\frac{1}{m}+\gamma\frac{i|\omega| \sqrt{\bar \rho}}{ |q|^3},
\label{eq:Pi0Pi2}
\end{equation}
where $\bar \rho$ is the electron density, $m$ is the electron bare (band) mass, and $\gamma = 2 \sqrt{3}$ is a numerical constant. Here, we have temporarily switched off the external probe electromagnetic gauge field $\delta A_\mu$ for clarity of the presentation. Here the momentum $q =|\bm{q}|$ carried by the gauge field is much smaller than the Fermi momentum $k_F$ of the composite fermions, i.e., $q \ll k_F$, and $\omega\ll v_Fq $, where the quantum critical dynamics is manifest. 

In contrast to the incompressible FQH states, where the Chern-Simons term is the most relevant term and dominates the low-energy physics, in the CFL the damping term $\propto  i\frac{\omega}{q^3}$ is the most relevant term in the low-energy regime. Furthermore, by gauge invariance, the gauge fields must remains gapless, which hence also play a key role in the low-energy physics of the nematic phase transition inside the composite Fermi liquid state of section \ref{sec:Nematic-CFL}. 

By restoring the probe field $\delta A_\mu$ and integrating out the low-energy fluctuation $\delta a_\mu$ in Eq.\eqref{effectiveCFL}, we can also derive effective action for the probe external  field $\delta A_\mu$, which encodes de correlation functions of the densities and currents of the CFL,
\begin{equation}
\mathcal{S}_{\rm eff}[\delta A_\mu]=-\int_{\bm{q},\omega} \frac{1}{2}\delta A_{\mu}(\bm{q},\omega) K_{\mu \nu}(\bm{q},\omega) \delta A_{\nu}(-\bm{q},-\omega) 
\end{equation}
where $K_{\mu\nu}(\bm{q},\omega)$ is the Fourier transform of the polarization tensor of the CFL. Its components are given by\cite{Lopez-1993,Halperin-1993}
\begin{align}
K_{00}(\bm{q},\omega)=&\frac{1}{2\pi}\bm{q}^2 K_0(\bm{q},\omega), \nonumber \\
K_{0i}(\bm{q},\omega)=&\frac{1}{2\pi}(\omega q_i K_0(\bm{q},\omega)+i \epsilon_{ik} q_k K_1(\bm{q},\omega)), \nonumber\\
K_{i0}(\bm{q},\omega)=&\frac{1}{2\pi}(\omega q_i K_0(\bm{q},\omega)-i \epsilon_{ik} q_k K_1(\bm{q},\omega)), \nonumber\\
K_{ij}(\bm{q},\omega)=&\frac{1}{2\pi}(\omega^2 \delta_{ij} K_0(\bm{q},\omega)-i\epsilon_{ij} \omega K_1(\bm{q},\omega)\nonumber\\
          &+(\bm{q} ^2 \delta_{ij}-q_iq_j)K_2(\bm{q},\omega)), 
          \label{eq:Kmunu}
\end{align}
where  the functions $K_0(\bm{q},\omega)$,  $K_1(\bm{q},\omega)$ and  $K_2(\bm{q},\omega)$  are given by 
\begin{align}
K_0(\bm{q},\omega)=&-\frac{\Pi_0(\bm{q},\omega)}{4D(\bm{q},\omega)}, \nonumber\\
K_1(\bm{q},\omega)=&\frac{1}{2}+\frac{1}{8 D(\bm{q},\omega)}+\frac{V({\bm q})\Pi_0(\bm{q},\omega) {\bm q}^2}{16\pi^2 D(\bm{q},\omega)}, \nonumber\\
K_2(\bm{q},\omega)=&\frac{\Pi_2(\bm{q},\omega)}{4 D(\bm{q},\omega)}+\frac{V({\bm q})\omega^2\Pi_0(\bm{q},\omega)^2}{ 16\pi^2  D(\bm{q},\omega)}\nonumber\\
&+\frac{V(q)\Pi_0(\bm{q},\omega) \Pi_2(\bm{q},\omega) {\bm q}^2}{ 16\pi^2 D(\bm{q},\omega)}, \nonumber\\
D(\bm{q},\omega)=&\Pi_0(\bm{q},\omega)^2 \omega^2-(\frac{1}{2})^2\nonumber\\
&+\Pi_0(\bm{q},\omega)(\Pi_2(\bm{q},\omega)-\frac{V(\bm {q})}{16\pi^2}){\bm q}^2.
\label{polEM}
\end{align}
From these expressions, we can calculate the spectrum of collective modes. In the incompressible FQH state the lowest energy collective mode is the  Girvin-MacDonald-Platzman (GMP), which in Ref.[\onlinecite{you2014}] we showed condenses at the nematic quantum phase transition. From  the polarization tensor of the external electromagnetic field, Eq.\eqref{eq:Kmunu}. we can extract the analog of the GMP mode for the HLR state.
The pole in the polarization tensor component $K_{00}(\bm{q},\omega)$ determines the collective mode. In the limit of $q\ll k_F$ and $\omega\ll v_Fq$, the gapless collective excitation of the half-filled Landau level is, 
\begin{align}
&\omega_{\pm}\sim \frac{i\sqrt{3}}{|q|} \Big[1  \pm \Big(1- \frac{|q|^3}{12\pi^2 k_F^3}(1+mV(\bm{q}))\Big)^{1/2} \Big]
\end{align}
From the residue of the polarization tensor for this mode we find a structure factor $S(q) \propto |\bm{q}|^4$, in analogy of the Girvin-MacDonald-Platzman mode in FQHE states.

Now we turn our attention to the composite fermions, and calculate the self-energy correction $\Sigma_f(\bm{k},\omega)$ for the composite fermion propagator by calculating the one-loop diagram shown in Fig. 5. The inverse fermion propagator $g^{-1}(\bm{k},\omega)$ is
\begin{equation}
g^{-1}(\bm{k}, \omega)=g_{0}^{-1}(\bm{k}, \omega) - \Sigma_f (\bm{k},\omega)
\end{equation}
where $g_0(\bm{k},\omega)$ is the free fermion propagator. The imaginary part of the self-energy $\Sigma''(\bm{k},\omega)$ is 
\begin{equation}
\Sigma''(\bm{q},\omega)\simeq -2\sqrt{3} ~ \;  \sign(\omega) (\frac{\beta |\omega|}{4\pi })^{2/3}
\label{FermionSelfE}
\end{equation}
where $\beta=\frac{k_F}{4\sqrt{m}}+V(0)k_F\sqrt{m}$, and where  we have dropped the bare term $\omega$ which is much smaller than the correction $|\omega|^{2/3}$.  Here we have assumed the density-density interaction is short-ranged, and hence only $V(\bm{q}\approx 0)$ appears in the expression. Here and after, we only consider the short-ranged density-density interaction.  This result shows that the singular forward scattering interaction with the  fluctuating gauge boson softens the Fermi surface and that the composite fermion quasiparticle is no longer well defined. The gauge boson drives the composite fermion into a non Fermi liquid.\cite{Halperin-1993}

\begin{figure} [h!] 
\begin{center}  
\includegraphics[width=0.3\textwidth]{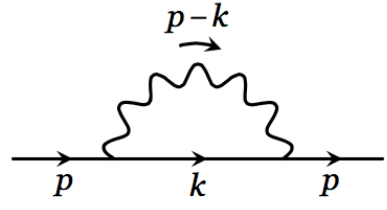}  
\caption{The gauge boson correction to the fermion propagator.}
\end{center}
\label{fig:self-energy}  
\end{figure}

 
\section{Coupling between the Gauge field and nematic field}
\label{app:coupling-gauge}

We start from the theory of the Fermi surface coupled to the Chern-Simons gauge field,
\begin{widetext} 
\begin{align}
\mathcal{S}=&\int d^2 r dt~ \Psi^{\dagger}(\bm{r},t)[D_t+\frac{1}{2m}\bm{D}^2+\frac{M_1}{2m}(D_x^2-D_y^2)+\frac{M_2}{2m}(2D_xD_y)]\Psi(\bm{r},t)\nonumber\\
&+\frac{1}{8\pi}\int d^2 r dt~\epsilon^{\mu \nu \rho} \delta a_{\mu} \partial_{\nu} \delta a_{\rho} +\int d^2rd^2r' dt V(\bm{r-r'})\delta \rho(\bm{r},t) \delta \rho(\bm{r'},t)
\end{align}
\end{widetext}
where $D_\mu=\partial_\mu+\delta a_\mu$.
The external magnetic field is screened by the Chern-Simons flux while $\delta a_i$ represent the gauge fluctuation.
In this mean-field level, the average flux felt by the composite fermion is zero so the CF is in the Fermi liquid state. If we consider gauge fluctuation to the quadratic order, we obtain the effective polarization for the gauge field
\begin{equation}
\mathcal{S}_{a}=-\frac{1}{2}\int_{\bm{q},\omega}\delta a_{\mu}(\bm{q},\omega) \Pi_{\mu\nu}(\bm{q},\omega) \delta a_{\nu}(\bm{-q},-\omega)
\end{equation}
where
\begin{widetext}
\begin{equation}
\Pi_{\mu \nu}=\frac{1}{2\pi}
\begin{pmatrix} 
q^2\Pi_0 & q_x\omega\Pi_0+iq_y/2 & q_y\omega\Pi_0 -iq_x/2\\
q_x\omega\Pi_0-iq_y/2 & \omega^2\Pi_0-q_y^2(\Pi_2+V(\bm{q}))  & q_yq_x(\Pi_2+V(\bm{q}))-i\omega/2\\
q_y\omega\Pi_0+iq_x/2 & q_yq_x(\Pi_2+V(\bm{q}))+i\omega/2 &\omega^2\Pi_0-q_x^2(\Pi_2+V(\bm{q}))
\end{pmatrix}
\label{eq:Pimunu}
\end{equation}
\end{widetext}
with
\begin{align}
\Pi_0(\bm{q},\omega) &=\frac{m}{q^2}-\frac{i|\omega| }{ \sqrt{{\bar \rho}} |q|^3},\label{eq:Pi0}\\
\Pi_2(\bm{q},\omega) &=\frac{1}{m}+\gamma\frac{i|\omega| \sqrt{{\bar \rho}}}{ |q|^3},\label{eq:Pi2}
\end{align}
where ${\bar \rho}$ is the electron density. 

The linear, parity-even,  coupling between the nematic field and the gauge field, i.e. the quadrupolar coupling between nematic fluctuations and gauge fields,  at mean field level has the form,
\begin{equation}
\mathcal{S}_{a,M}[\bm{M},\delta a_\mu]=-\frac{1}{2}\int_{\bm{q},\omega} M_i (\bm{q},\omega) T_{i \nu}(\bm{q},\omega) \delta a_{\nu}(\bm{q},\omega)
\label{eq:SaM}
\end{equation}
where
\begin{align}
T(\bm{q},\omega)=&\frac{\Pi_2(\bm{q},\omega) m}{2\pi}
\begin{pmatrix} 
q_x^2-q_y^2 & q_x\omega & -q_y\omega\\
2q_xq_y & \omega q_y  & q_x \omega
\end{pmatrix},\nonumber\\
\Pi_2(\bm{q},\omega) =&\gamma\frac{i|\omega| \sqrt{{\bar \rho}}}{ |q|^3}
\end{align}
where we used the overdamped form of $\Pi_2(\bm{q},\omega)$,   for $1 \gg \frac{\omega}{qv_F}\gg q^2$.

We also have the coupling  between the nematic order parameter and the Chern-Simons gauge fields, 
 \begin{equation}
\mathcal{S}_{a,M,a}[\delta a_\mu,\bm{M}]=-\frac{1}{2}\int_{\bm{q},\omega}\delta a_{\mu}(\bm{q},\omega) V_{\mu \nu}[\bm{M}] \delta a_{\nu}(\bm{-q},-\omega)
\label{eq:SaMa}
\end{equation}
\begin{widetext}
\begin{equation}
V_{\mu \nu}[\bm{M}]=\frac{\Pi_0(\bm{q},\omega)}{2\pi}
\begin{pmatrix} 
\frac{M_1}{2}(q_x^2-q_y^2) & (M_1 q_x+M_2 q_y)\omega  & (-M_2 q_y+M_2 q_x)\omega  \\
(M_1 q_x+M_2 q_y)\omega & M_1\omega^2  & M_2\omega^2 \\
(-M_2 q_y+M_2 q_x)\omega & M_2 \omega^2 & -M_1\omega^2 
\end{pmatrix}
\end{equation}
\end{widetext}  
This parity-even term represents the coupling of the gauge fields to the nematic fluctuations as a local fluctuation of the spatial metric.

\section{Details of calculation of the Berry phase term}
\label{app:details-berry}

In this section, we show the detailed calculation of the Berry phase term whose Feynman diagram is shown in Fig.2. 
We  obtained a similar term in the case of the FQH  in Ref.[\onlinecite{you2014}]. 
The Berry phase term is  obtained once we integrate out the gauge fluctuation and expand the theory in quadratic level. Here, we start from the gauge-nematic order parameter theory. Here we choose the temporal gauge $a_0=0$, and set $m=1$ (which we will restore back later).

\begin{widetext}
\begin{align}
&\mathcal{S}=-\int_{\bm{q},\omega} \frac{1}{2} \delta a_{i} (\bm{q},\omega)[\Pi_{ij} +t^1_{ij}( \omega^2 \Pi_0)  M_1+t^2_{ij} (\omega^2 \Pi_0) M_2)]\delta a_{j}(\bm{-q},-\omega)\nonumber\\
&\Pi^{-1} =\frac{\begin{pmatrix} 
 \omega^2 \Pi_0- q_x^2 \Pi_2 & -q_x q_y \Pi_2-i\omega/2 \\
-q_x q_y \Pi_2+i\omega/2  &  \omega^2 \Pi_0- q_y^2 \Pi_2
 \end{pmatrix}}{\omega^4 \Pi_0^2-q^2\omega^2 \Pi_0 \Pi_2-\omega^2/4}
\end{align}
\end{widetext}

Since we are looking at the limit $|\frac{\omega}{v_Fq}|<1$, we only keep the order of $O(\frac{\omega}{v_Fq})$ for the damping part in the gauge boson propagator and drop terms like $(\frac{\omega}{v_Fq})^2$. Afterwards, the inverse of the polarization tensor is,
\begin{align}
&\Pi^{-1} =-\frac{\begin{pmatrix} 
- q_x^2 \Pi_2 & -q_x q_y \Pi_2-i\omega/2 \\
-q_x q_y \Pi_2+i\omega/2  &  - q_y^2 \Pi_2
 \end{pmatrix}}{q^2\omega^2 \Pi_0 \Pi_2+\omega^2/4}
 \nonumber\\
 &\Pi_0(\bm{q},\omega) =\frac{m}{q^2},~~\Pi_2(\bm{q},\omega) =\gamma\frac{i|\omega| \sqrt{{\bar \rho}}}{ |q|^3},
\end{align}

Here we temporarily use  units in which $m=1$ for  convenience. Then the effective theory for the nematic order parameters can be obtained as
\begin{align}
&(-i)\frac{\delta \mathcal{S}}{\delta M_i \delta  M_j}(\Omega,\bm{Q})= \nonumber\\
&i \textrm{Tr} \int_{\bm{q},\omega}~\Pi ^{-1}(\omega,\bm{q}) \omega^2 \Pi_0 t_{i}  \Pi^{-1}(\omega+\Omega,\bm{q+Q}) \omega^2 \Pi_0 t_{j}\nonumber\\
&=i\Omega/2 \epsilon^{ij}~\int_{\bm{q},\omega}~\frac{\Pi_2 q^2 }{\Pi_2 q^2+q^2/4} \frac{1}{\Pi_2 q^2+q^2/4} 
\end{align}
where $t_1=\sigma_3$ and $t_2=\sigma_1$.

Thus the anti-symmetric part of the effective theory of the nematic order parameters is
\begin{equation}
\mathcal{L}^{M}_{ij}=\chi \epsilon^{ij} M_i(\Omega,\bm{Q})(i\Omega ) M_j(-\Omega,-\bm{Q})
\end{equation}
where
$\chi= \frac{2}{3\pi}  \bar{\Lambda}$, and
$\bar \Lambda$ is the frequency cut off. After putting back in the electron mass, we have $\chi= \frac{2}{3\pi}  \bar{\Lambda}m$.

\section{Details of calculations of the Wen-Zee term}
\label{app:details-WZ}

First, we calculate the parity-even part of the Wen-Zee term where the diagram has one gauge leg and one nematic leg (see Fig.\ref{fig:parity-even-mixing})
We use the temporal gauge $a_0=0$ and take  units in which $m=1$, and find
\begin{equation}
(-i)\frac{\delta \mathcal{S}}{\delta M_l \delta  a_k}= \langle  a_{m}  \omega^2 \Pi_0 t^{mn}_{l} a_n a_m  \omega^2 \Pi_0 Q^{mi}_{k} M_{i}  \rangle  
\end{equation}
where
\begin{equation}
T^{-1}_{mi}=\langle a_m M_{i} \rangle(\bm{q},\omega)=\frac{
\begin{pmatrix} 
 q_x\omega & -q_y\omega\\
 \omega q_y  & q_x \omega
\end{pmatrix}_{mi}}
{\Pi_2   q^2 \omega^2},
\end{equation}
with the notation
$Q_x= I ,Q_y= i \sigma_2$, and 
\begin{widetext}
\begin{align}
-i\frac{\delta \mathcal{S}}{\delta M_l \delta  a_k}(\bm{p},\Omega)&=i  \int _{\bm{q},\omega}~ \textrm{Tr}\Big[(\omega^2 \Pi_0)^2  \Pi^{-1}(\omega-\Omega/2,\bm{q-Q/2})  t_{l}  T^{-1}(\omega+\Omega/2,\bm{q+Q/2})  Q_{k}\Big] \nonumber\\
&=\begin{pmatrix} 
 p_x\Omega & -p_y\Omega\\
\Omega p_y  & p_x \Omega
\end{pmatrix}_{lk}
 \int_{\bm{p},\omega}  \frac{1}{2 q(\Pi_2 q^3+q^3/4)}  \nonumber\\
&=\frac{1}{6 \pi \sqrt{3}}(\frac{\bar{\Lambda}}{{\bar \rho}})^{1/3} \begin{pmatrix} 
 p_x\Omega & -p_y\Omega\\
 \Omega p_y  & p_x \Omega
\end{pmatrix}_{lk}
\end{align}
Restoring the mass back, we have
\begin{equation}
-i\frac{\delta \mathcal{S}}{\delta M_l \delta  a_k}(\bm{p},\Omega)= 
\frac{1}{6\pi \sqrt{3}}(\frac{\bar{\Lambda}m}{{\bar \rho}})^{1/3} 
\begin{pmatrix} 
 p_x\Omega & -p_y\Omega\\
 \Omega p_y  & p_x \Omega
\end{pmatrix}_{lk}
\end{equation}

To calculate the cubic level of the Wen-Zee term where the diagram has one gauge field external leg and two nematic fields external legs 
(shown in Fig. 4).
We will choose the axial gauge where $a_0=0$ (and use units which the effective mass is $m$=1), to find
\begin{equation}
-i\frac{\delta \mathcal{S}}{\delta M_h \delta M_l \delta  a_k}= 
 -\langle  a_{m}  \omega^2 \Pi_0 \sigma^{mn}_{h} a_n a_{m}  \omega^2 \Pi_0 \sigma^{mn}_{l} a_n a_m  \omega^2 \Pi_0 Q^{mi}_{k} M_{i}  \rangle
\end{equation}
Similarly,
\begin{align}
&(-i)\frac{\delta \mathcal{S}}{\delta M_h \delta M_l \delta  a_k}(\Omega_1,\Omega_2; \bm{p_1},\bm{p_2})\nonumber\\
&=- \int_{\bm{q} ,\omega}~ \textrm{Tr} \Big[\Pi^{-1}(\omega,\bm{q}) (\omega^2 \Pi_0 \sigma_{h}) \Pi^{-1}(\omega+\Omega_1,\bm{q+p_1}) (\omega^2 \Pi_0 \sigma_{l})  
T^{-1}(\omega+\Omega_1+\Omega_2,\bm{q+p_1+p_2}) (\omega^2 \Pi_0 Q_{k}) \Big]\nonumber\\
&=\epsilon^{hl} \epsilon^{\nu\mu k} p_1^{\nu}(p_1^{\mu}+p_2^{\mu})\int_{\bm{q} ,\omega}~ \Big[\frac{ |\omega| q^2}{8 q(\Pi_2 q^3+q^3/4)^2 }
+\frac{ \Pi_2 q^3}{8 q(\Pi_2 q^3+q^3/4)^2}\Big]\nonumber\\
&=\frac{1}{6 \pi \sqrt{3}}\Big[\left(\frac{\bar{\Lambda}}{{\bar \rho}}\right)^{1/3} +\frac{\bar{\Lambda}}{{\bar \rho}}\Big]\epsilon^{hl} \epsilon^{\nu\mu k} p_1^{\nu}(p_1^{\mu}+p_2^{\mu})
\end{align}
\end{widetext}
Restoring the mass back, we find the effective theory as
\begin{align}
&(-i)\frac{\delta \mathcal{S}}{\delta M_h \delta M_l \delta  a_k}(\Omega_1,\Omega_2; \bm{p_1},\bm{p_2})\nonumber\\
&=\frac{1}{6  \pi \sqrt{3}}\Big[\left(\frac{\bar{\Lambda}m}{{\bar \rho}}\right)^{1/3} +\frac{\bar{\Lambda}m}{{\bar \rho}}\Big]\epsilon^{hl} \epsilon^{\nu\mu k} p_1^{\nu}(p_1^{\mu}+p_2^{\mu})
\end{align}

Similar to the calculation of the Berry phase term by Ward Identity, the Wen-Zee term can also be derived from the Ward Identity.
The nematic order parameter modifies the local geometry metric of the Maxwell term and couples to the bilinear of the Chern-Simons gauge fields. At this level, we can integrate out the gauge fields and perform the loop expansion of the nematic fields at the cubic level.
\begin{widetext}
\begin{align}
\langle T_1 T_2 T_{xy}\rangle &(\bm{p}_1,\bm{p}_2,\Omega_1,\Omega_2)=\nonumber\\
&-\int_{\bm{k}, \omega}\textrm{Tr}\Big[ \Pi^{-1}(\bm{k},\omega) (\Pi_0 \omega^2) \sigma_{z} \Pi^{-1}(\bm{k+p_1},\omega+\Omega_1)  (\Pi_0 \omega^2) \sigma_{z} \Pi^{-1}(\bm{k+p_1+p_2},\omega+\Omega_1)  (\Pi_0 \omega^2)\sigma_{+}\Big]
\end{align}
By taking advantage of the Ward Identity, we find
\begin{align}
m \partial_0  J_x+\partial_{x}T_{xx}+\partial_{y}T_{yx}=\delta b  \cdot  J_y \nonumber\\
 m \partial_0  J_y+\partial_{x}T_{xy}+\partial_{y}T_{yy}=-\delta b \cdot  J_x 
\end{align}
We can set up several relations between the Wen-Zee term and the terms cubic in nematic order parameters,
\begin{align}
m \langle T_1(\bm{r}_1) T_2(\bm{r}_2) \partial_0 J_x (\bm{r}_3)\rangle+\langle T_1(\bm{r}_1) T_2(\bm{r}_2) \partial_{x}T_{xx}(\bm{r}_3)\rangle+\langle T_1(\bm{r}_1) T_2(\bm{r}_2) \partial_{y}T_{yx}(\bm{r}_3)\rangle=2 \langle T_1(\bm{r}_1) T_2(\bm{r}_2) \delta \rho(\bm{r}_3)  J_y (\bm{r}_3) \rangle\nonumber\\
m \langle T_1(\bm{r}_1) T_2(\bm{r}_2) \partial_0 J_y (\bm{r}_3)\rangle+\langle T_1(\bm{r}_1) T_2(\bm{r}_2) \partial_{x}T_{xy}(\bm{r}_3)\rangle+\langle T_1(\bm{r}_1) T_2(\bm{r}_2) \partial_{y}T_{yy}(\bm{r}_3)\rangle=-2 \langle T_1(\bm{r}_1) T_2(\bm{r}_2) \delta \rho(\bm{r}_3)  J_x (\bm{r}_3) \rangle
\end{align}
It is straightforward to see that the correlators on the right hand sides are actually zero. Also, the further calculation shows that $\langle T_1(\bm{r}_1) T_2(\bm{r}_2) \partial_{x}T_{xx}(\bm{r}_3)\rangle$ and $\langle T_1(\bm{r}_1) T_2(\bm{r}_2) \partial_{y}T_{yy}(\bm{r}_3)\rangle$ does not generate any anti-symmetric term at the leading order ($O(q^3)$).
Now the identities become,
\begin{align}
m \langle T_1(\bm{r}_1) T_2(\bm{r}_2) \partial_0 J_x (\bm{r}_3)\rangle=-\langle T_1(\bm{r}_1) T_2(\bm{r}_2) \partial_{y}T_{yx}(\bm{r}_3)\rangle\nonumber\\
m \langle T_1(\bm{r}_1) T_2(\bm{r}_2) \partial_0 J_y (\bm{r}_3)\rangle=-\langle T_1(\bm{r}_1) T_2(\bm{r}_2) \partial_{x}T_{xy}(\bm{r}_3)\rangle
\end{align}
We have
\begin{align}
\langle T_i  T_j \partial_{y}T_{yx}\rangle(\bm{p_1,p_2},\omega_1,\omega_2)=s  \epsilon^{ij} \omega_1 (\omega_1+\omega_2) (-p^y_1-p^y_2)-s \epsilon^{ij} \omega_1 (\omega_1+\omega_2) p^y_1   \nonumber\\
\langle T_i T_j \partial_{x}T_{xy}\rangle(\bm{p_1,p_2},\omega_1,\omega_2)=-s  \epsilon^{ij} \omega_1 (\omega_1+\omega_2) (-p^x_1-p^x_2)+s \epsilon^{ij} \omega_1 (\omega_1+\omega_2) p^x_1 
\end{align}
Thus, the Wen-Zee term has the form
\begin{align}
\langle T_i  T_j J_x\rangle(\bm{p_1,p_2},\omega_1,\omega_2)=\frac{s}{m} \epsilon^{ij} \omega_1  (-p^y_1-p^y_2)-s \epsilon^{ij} (-\omega_1-\omega_2)  p^y_1 \nonumber\\
\langle T_i  T_j J_y\rangle(\bm{p_1,p_2},\omega_1,\omega_2)=-\frac{s}{m} \epsilon^{ij} \omega_1  (-p^x_1-p^x_2)+s \epsilon^{ij} (-\omega_1-\omega_2)  p^x_1
\end{align}
\end{widetext}
By taking advantage of the current conservation relation,
\begin{align}
\partial_{\mu} J_{\mu}=0
\end{align}
We obtain
\begin{align}
\langle T_i  T_j J_0\rangle(\bm{p_1,p_2},\omega_1,\omega_2)=&\frac{s}{m} \epsilon^{ij} p^y_1  (-p^x_1-p^x_2)\nonumber\\
-&s \epsilon^{ij} (-p^y_1-p^y_2)  p^x_1
\end{align}
Where the coefficient $\frac{s}{m}$ is obtained to be
\begin{align}
\frac{s}{m}=\frac{1}{6 \pi \sqrt{3}}\Big[\left(\frac{\bar{\Lambda}m}{{\bar \rho}}\right)^{1/3} +\frac{\bar{\Lambda}m}{{\bar \rho}}\Big]
\end{align}
Thus, we finally obtain that the Wen-Zee term $\mathcal{L}_{WZ}$ of the effective Lagrangian is
\begin{align}
\mathcal{L}_{WZ}=\frac{1}{6 \pi \sqrt{3}}\Big[\left(\frac{\bar{\Lambda}m}{{\bar \rho}}\right)^{1/3} +\frac{\bar{\Lambda}m}{{\bar \rho}}\Big]\ ~\epsilon^{\mu\nu\rho}\omega_{\mu}^{Q} \partial_{\nu}  A_{\rho}
\end{align}

\section{Nematic correlators}
\label{app:nematic-correlators}

The self-energy of the composite fermion close to the Fermi surface is,
\begin{align}
&\Sigma_s(\omega)=- i2\sqrt{3}  \;  \sign(\omega) (\frac{\beta \omega}{4\pi })^{2/3}
\end{align}
The Fermi surface is softened by the gauge boson and thus no longer well defined. Here we check the correction from the self-energy to of the nematic correlator.

The nematic correlator of the Fermi surface without self-energy correction is shown in the Feynman diagram of Fig.\ref{NCorrelator}, and it is given by
\begin{widetext}
\begin{align}
-i\frac{\delta S}{\delta M_1 \delta M_1 }&=\langle T_1 T_1 \rangle_0 (\bm{p},\Omega) \nonumber\\
&=-i \int_{\bm{k}, \omega}~ g (\bm{k+p},\omega+\Omega) \frac{(k_x^2-k_y^2)}{2m} g (\bm{k},\omega) \frac{(k_x^2-k_y^2)}{2m} \nonumber\\
&=-i \int_{\bm{k}, \omega}~\frac{1}{4m^2}\frac{(k_x^2-k_y^2)}{\omega+\Omega-((\bm{k+p})^2+k_F^2)/2m
+i\eta  \;  \sign(\omega+\Omega)}~\frac{(k_x^2-k_y^2)}{\omega-(\bm{k}^2+k_F^2)/2m+i\eta  \;  \sign(\omega)} \nonumber\\
&=k_F^4/(4\pi m)-\cos^2(2 \theta_p)\frac{k_F^3}{2\pi} (i\Omega/p)-\sin^2(2 \theta_p)\frac{k_F^2 m}{2\pi}(\Omega/p)^2
\end{align}

If we only look into the leading self-energy correction, it is the sum of the diagrams in Fig. \ref{nematiconetwo},
\begin{figure}  
\begin{center}
\subfigure[]{\includegraphics[width=0.35\textwidth]{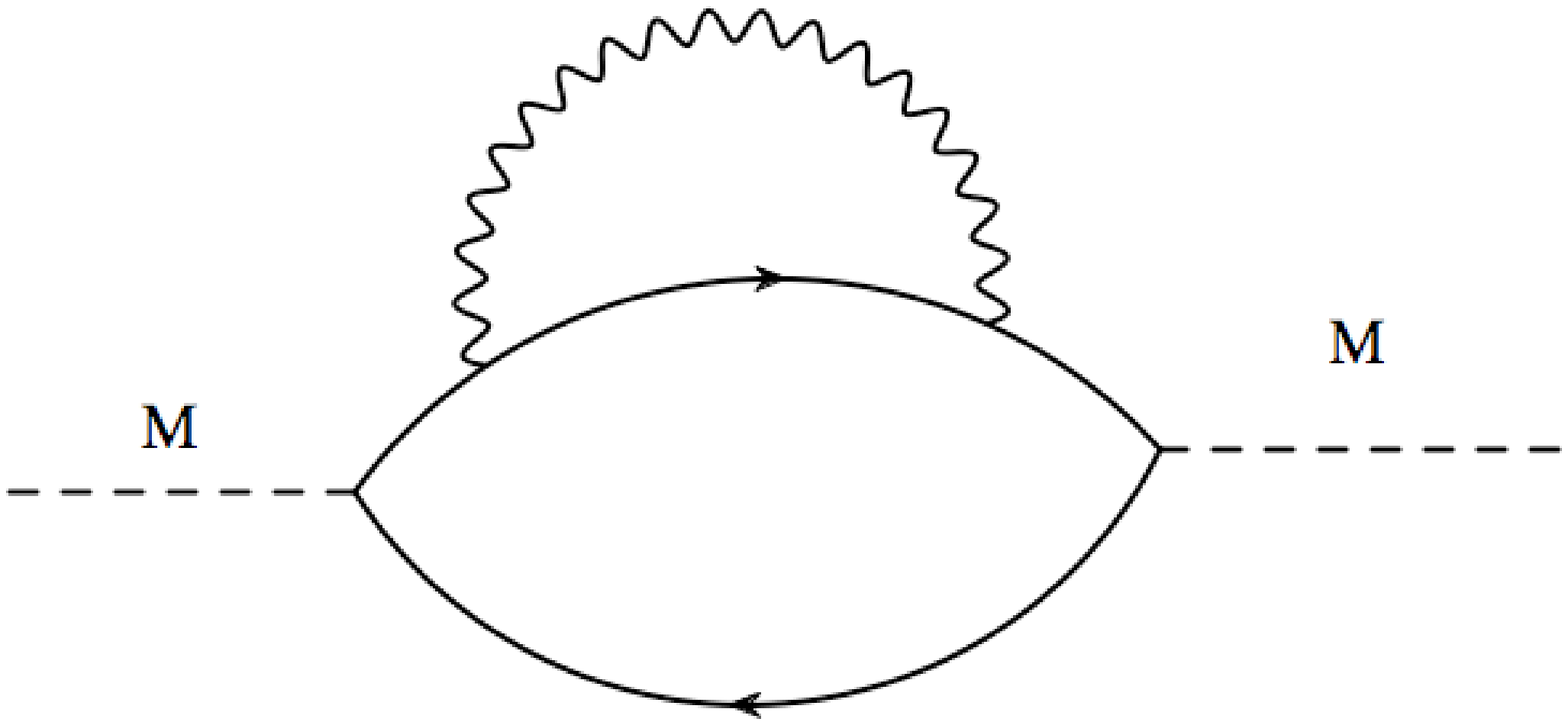}}
\subfigure[]{\includegraphics[width=0.35\textwidth]{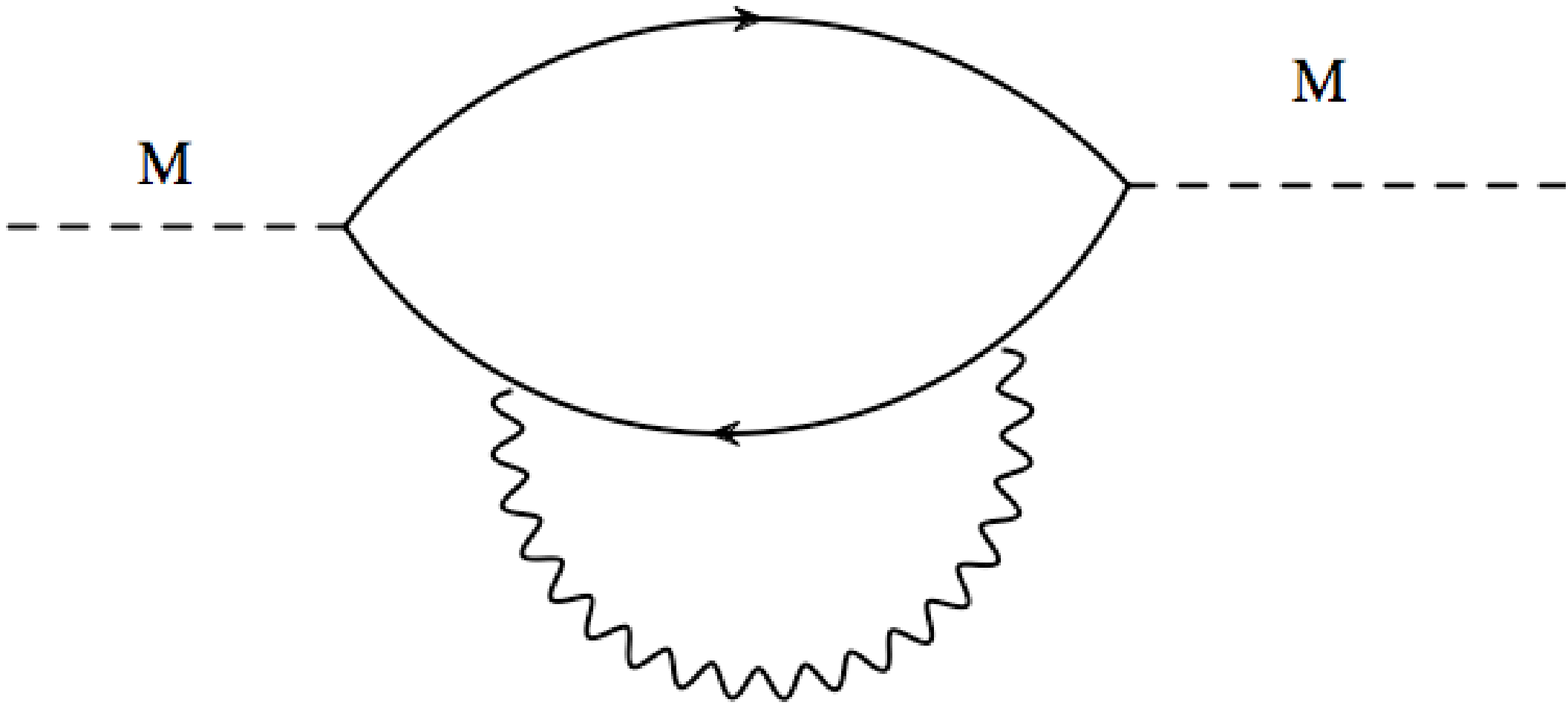}}
\subfigure[]{\includegraphics[width=0.35\textwidth]{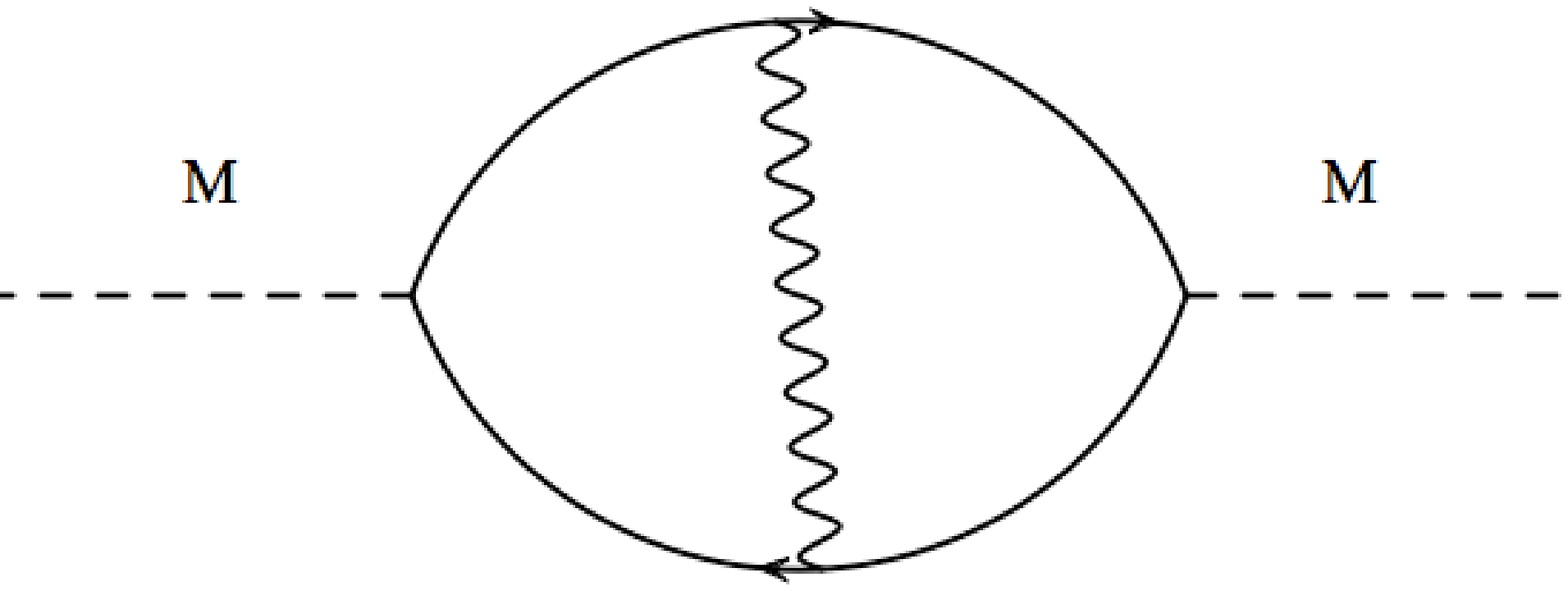}}
\end{center}
\caption{Leading corrections from gauge field fluctuations to the nematic correlators: a) and b)  composite fermion self-energy corrections, and c) vertex correction.}
\label{nematiconetwo}
\end{figure}
\begin{align}
\langle T_1 T_1 \rangle_1 (\bm{p},\Omega)&= \nonumber\\
&-i \int_{\bm{k}, \omega}~ (k_x^2-k_y^2)^2/(4m^2)[g (\bm{k+p},\omega+\Omega) \Sigma_s(\omega) g^2 (\bm{k},\omega) +g^2 (\bm{k+p},\omega+\Omega) \Sigma_s(\omega+\Omega) g (\bm{k},\omega)] 
\end{align}
Here the self-energy $\Sigma_s(\omega)$ only affects the fermions near the Fermi surface where $k\sim k_F$,
Take advantage of the identity
\begin{align}
&g (\bm{k+p},\omega+\Omega) \Sigma_s(\omega) g^2 (\bm{k},\omega)=-\frac{\Sigma_s(\omega)}{\Omega-\bm{k}\cdot \bm{p}/2m}[g^2 (\bm{k},\omega)-g (\bm{k+p},\omega+\Omega) g (\bm{k},\omega)] \nonumber\\
&g^2 (\bm{k+p},\omega+\Omega) \Sigma_s(\omega+\Omega) g (\bm{k},\omega)=\frac{\Sigma_s(\omega+\Omega)}{\Omega-\bm{k}\cdot \bm{p}/2m}[g^2 (\bm{k+p},\omega+\Omega)-g (\bm{k+p},\omega+\Omega)g (\bm{k},\omega)]
\end{align}
We can rewrite the integral as
\begin{equation}
\langle T_1 T_1 \rangle_1 (\bm{p},\Omega)=
-i \int_{\bm{k}, \omega}~ (k_x^2-k_y^2)^2/(4m^2) \frac{\Sigma_s(\omega)-\Sigma_s(\omega+\Omega)}{\Omega-\bm{k} \cdot \bm{p}/2m} g (\bm{k+p},\omega+\Omega)  g (\bm{k},\omega)
\end{equation}
In the same way, the other nematic correlators can be written as
\begin{align}
\langle T_2 T_2 \rangle_1 (\bm{p},\Omega)&=
-i \int_{\bm{k}, \omega}~ (2k_x k_y)^2/(4m^2)  \frac{\Sigma_s(\omega)-\Sigma_s(\omega+\Omega)}{\Omega-\bm{k} \cdot \bm{p}/2m}  g (\bm{k+p},\omega+\Omega)  g (\bm{k},\omega)  \nonumber\\
\langle T_2 T_1 \rangle_1 (\bm{p},\Omega)&=
-i \int_{\bm{k}, \omega}~ (k_x^2-k_y^2)(2k_x k_y)/(4m^2)  \frac{\Sigma_s(\omega)-\Sigma_s(\omega+\Omega)}{\Omega-\bm{k} \cdot \bm{p}/2m}  g (\bm{k+p},\omega+\Omega)  g (\bm{k},\omega)  
\end{align}
Here we first calculate  $\langle T_1 T_1 \rangle_1$,
\begin{align}
\langle T_1 T_1 \rangle_1 &(\bm{p},\Omega)= \nonumber\\
=-i \int_{\bm{k}, \omega}~ &\frac{k^4 \cos^2(2\theta)}{4m^2} \frac{\Sigma_s(\omega)-\Sigma_s(\omega+\Omega)}{\Omega-kp\cos(\theta-\theta_p)/2m} \frac{1}{\omega+\Omega-(E(k)+kp\cos(\theta-\theta_p)/2m)+i\eta  \;  \sign(\omega+\Omega)} \frac{1}{\omega-E(k)+i\eta  \;  \sign(\omega)}\nonumber\\
\end{align}
where $E(k)=(k^2-k_F^2)/2m$.

Since the self energy correction $\Sigma(\omega)$ only appears near the Fermi surface, we can replace $k/2m=v_F=k_F/2m$ in the calculation to simplified the problem. We can replace the integral $d\bm{k}=dE(k) d \theta~2m$,
\begin{align}
&\langle T_1 T_1 \rangle_1 (\bm{p},\Omega)= \nonumber\\
&-i \int d\omega dE(k) d \theta ~\frac{k_F^4 \cos^2(2\theta)}{4m}  \frac{\Sigma_s(\omega)-\Sigma_s(\omega+\Omega)}{\Omega-v_F p\cos(\theta-\theta_p)} \frac{1}{\omega+\Omega-(E(k)+v_F p\cos(\theta-\theta_p))+i\eta  \;  \sign(\omega+\Omega)} \frac{1}{\omega-E(k)+i\eta  \;  \sign(\omega)}\nonumber\\
&= \int d\omega d \theta ~\frac{k_F^4 \cos^2(2\theta)}{4m} \frac{\Sigma_s(\omega)-\Sigma_s(\omega+\Omega)}{(\Omega-v_F p\cos(\theta-\theta_p))^2} ( \;  \sign(\omega)- \;  \sign(\omega+\Omega))
\end{align}
First, we look at the static limit $\Omega=0$. Then we have $\Sigma_s(\omega)-\Sigma_s(\omega+\Omega)=0$ and therefore the integral is zero. This indicates the leading order self-energy correction does not generate any constant term, and thus the mass term of the nematics remains unchanged by the imaginary part of self-energy. 

Now we turn to the case when $\Omega$ is nonzero and take the limit $p \ll k_F$ and $\Omega \ll p$. The integral of $\theta$ can be done by contour integral, we take the leading order in regard of $\Omega/p$. Since we are only interested in the leading order behavior of the damping term, we calculate $\langle (T_1 T_1+ T_2 T_2) \rangle_1 $ to simplify the calculation.
\begin{align}
\langle T_i T_i \rangle_1 (\bm{p},\Omega)&= \nonumber\\
&=\int d\omega d \theta ~\frac{k_F^4 }{4m}  \frac{\Sigma_s(\omega)-\Sigma_s(\omega+\Omega)}{(\Omega-v_F p\cos(\theta-\theta_p))^2} ( \;  \sign(\omega)- \;  \sign(\omega+\Omega))\nonumber\\
&= \int d\omega  ~\frac{k_F^4 }{4m} \frac{\Sigma_s(\omega)-\Sigma_s(\omega+\Omega)}{(v_F p)^2} ( \;  \sign(\omega)- \;  \sign(\omega+\Omega))\frac{\Omega}{v_Fp}\frac{1}{(1+(\Omega/v_Fp)^2)^{3/2}} \nonumber\\
& \approx  \int d\omega  ~\frac{k_F^4 }{4m} \frac{\Sigma_s(\omega)-\Sigma_s(\omega+\Omega)}{(v_F p)^2} ( \;  \sign(\omega)- \;  \sign(\omega+\Omega))\frac{\Omega}{v_Fp}\nonumber\\
&= ~ \frac{k_F^4 \sqrt{3}}{2m}\frac{\beta^{2/3} \Omega^{5/3}}{(v_F p)^2} \frac{\Omega}{v_Fp}
\end{align}
The leading order damping upon the self-energy correction is $(\Omega)^{8/3}/q^3$, which is highly irrelevant so we can ignore it. However, one might worry that $\langle T_1 T_1 \rangle_1$ could contain a singular damping term which could cancel when added to $\langle T_2 T_2 \rangle_1$. If this cancellation happened, then the damping coefficient $\langle T_1 T_1 \rangle_1$ would be a function of $\cos(4 \theta_p)$. We can then choose $ \theta_p=0$ and calculate $\langle T_1 T_1 \rangle_1$ to check the leading damping term. The leading order of damping for $\langle T_1 T_1 \rangle_1( \theta_p=0)$ is of order $\Omega^{5/3}/q^2$, which is still irrelevant compared to the damping term.

In conclusion, the leading order self-energy correction does not change the leading order behavior of the nematic transition. In sum, the nematic polarization tensor is 
\begin{align}
&\langle T_i T_j \rangle (\bm{p},\Omega)= \nonumber\\
&\begin{pmatrix} 
k_F^4/(4\pi m)-\cos^2(2 \theta_p)\frac{k_F^3}{2\pi} (i\Omega/p)-\sin^2(2 \theta_p)\frac{k_F^2 m}{2\pi}(\Omega/p)^2  &\sin(2 \theta_p)\cos(2 \theta_p)\frac{k_F^3}{2\pi} (i\Omega/p) \\ 
\sin(2 \theta_p)\cos(2 \theta_p)\frac{k_F^3}{2\pi} (i\Omega/p)) & k_F^4/(4\pi m)-\sin^2(2 \theta_p)\frac{k_F^3}{2\pi} (i\Omega/p)-\cos^2(2 \theta_p)\frac{k_F^2 m}{2\pi}(\Omega/p)^2
\end{pmatrix}
\end{align}

\section{Vertex correction for nematic polarization tensor}
\label{app:vertex}

In this section, we  show the vertex correction for the nematic polarization tensor is irrelevant in our theory (shown in Fig.\ref{nematiconetwo}c)

\begin{align}
&\langle T_1 T_1 \rangle_v (\bm{p},\Omega) +\langle T_2 T_2 \rangle_v (\bm{p},\Omega)= \nonumber\\
&-i \int_{\bm{k}, \bm{q},\omega,\nu}~ \frac{k^6}{4m^4} g (\bm{k+p}, \omega + \Omega)  g (\bm{k},\omega) g(\bm{k+p+q},\omega+\Omega+\nu) 
g (\bm{k+q}, \omega+ \Omega) D_{11}(\bm{q},\nu) 
\end{align}


To proceed, we first calculate the vertex,
 \begin{align}
\Gamma(\bm{k},\bm{p},\omega,\Omega)&=  \nonumber\\
&-i \int_{\bm{q},\nu}~ g(\bm{k+p+q},\omega+\Omega+\nu) g (\bm{k+q}, \omega+ \Omega) D_{11}(\bm{q},\nu)   \nonumber\\
&-i \int_{\bm{q},\nu}~\frac{|q_v|}{(\omega+\Omega+\nu-(E(k)+v_F q_l+v_F p \cos(\theta-\theta_p)+q_v p \sin(\theta-\theta_p))
+i\eta  \;  \sign(\omega+\Omega+\nu))} \nonumber\\
& \times \frac{1}{(a|\nu|+b|q_v| ^3)}~\frac{1}{\omega+\nu-(E(k)+v_F q_l)+i\eta  \;  \sign(\omega+\nu)} 
\end{align}
where $q_v=q\sin(\theta-\theta_q),q_l=q\cos(\theta-\theta_q)$. $a = 2m\sqrt{3} $ and $b = 1/4$ are constants which we will take as unity for simplicity since we are only interested in scaling behaviors. Taking the integral $q_l$ first, we find
 \begin{align}
\Gamma(\bm{k},\bm{p},\omega,\Omega) &= \nonumber\\
& -i\int_{q_v,\nu}~\frac{|q_v|( \;  \sign(\omega+\Omega+\nu)- \;  \sign(\omega+\nu))}{(\Omega-v_F p \cos(\theta-\theta_p)+q_v p \sin(\theta-\theta_p)
)(a|\nu|+b|q_v| ^3)} \nonumber\\
&\sim  -i\int_{q_v}~\frac{|q_v|}{\Omega-v_F p \cos(\theta-\theta_p)+q_v p \sin(\theta-\theta_p)
} \Big(\ln(1+\frac{|\omega+\Omega|}{|q_v|^3}) \;  \sign(\omega+\Omega)-\ln(1+\frac{|\omega|}{|q_v|^3}) \;  \sign(\omega)\Big)
\end{align}
It is obvious that when $\Omega=0$, this expression is vanishes. Thus the vertex correction does not contribute to the mass term of the polarization tensor. 

Now we turn to the damping term in the vertex corrections to the polarization tensor,
\begin{align}
\langle T_i T_i \rangle_v (\bm{p},\Omega) &= \nonumber\\
& -i\int_{q_v,\bm{k},\omega}~\frac{|q_v|}{\Omega-v_F p \cos(\theta-\theta_p)+q_v p \sin(\theta-\theta_p)
} \Big(\ln(1+\frac{|\omega+\Omega|}{|q_v|^3}) \;  \sign(\omega+\Omega)-\ln(1+\frac{|\omega|}{|q_v|^3}) \;  \sign(\omega)\Big)\nonumber\\
& \times \frac{k^6}{4m^4} \Big(\omega+\Omega-E(k)-v_F p \cos(\theta-\theta_p)+i\eta   \;  \sign(\omega+\Omega))^{-1}  (\omega-E(k)+i\eta   \;  \sign(\omega)\Big)^{-1}
\end{align}
As the gauge field correction only affects the fermion propagator near the Fermi surface, we can take $\frac{k^6}{4m^4}=\frac{k_F^6}{4m^4}$ and take this constant aside. We can then replace the integral over $d\bm{k}=dE(k) d \theta~2m$ and integrate over $dE(k) $ first,
\begin{align}
\langle T_i T_i \rangle_v (\bm{p},\Omega) \sim\int_{q_v,\theta,\omega}~&\frac{|q_v|}{\Omega-v_F p \cos(\theta-\theta_p)+q_v p \sin(\theta-\theta_p)
} \Big(\ln(1+\frac{|\omega+\Omega|}{|q_v|^3}) \;  \sign(\omega+\Omega)-\ln(1+\frac{|\omega|}{|q_v|^3}) \;  \sign(\omega)\Big)\nonumber\\
& \times \frac{ \;  \sign(\omega+\Omega)- \;  \sign(\omega)}{\Omega-v_F p \cos(\theta-\theta_p)}\nonumber\\
\sim \int_{q_v,\theta}~ \int_{0}^{\Omega} d\omega &~\frac{|q_v|}{\Omega-v_F p \cos(\theta-\theta_p)+q_v p \sin(\theta-\theta_p)} \ln(1+\frac{|\omega|}{|q_v|^3}) \frac{1}{\Omega-v_F p \cos(\theta-\theta_p)}
\end{align}
By the change of variables $l=\omega/\Omega$, we get
\begin{equation}
\langle T_i T_i \rangle_v (\bm{p},\Omega) \sim 
\int_{q_v,\theta}~ \int_{0}^{1} dl ~\frac{\Omega |q_v|}{\Omega-v_F p \cos(\theta-\theta_p)+q_v p \sin(\theta-\theta_p)} \ln(1+\Omega \frac{|l|}{|q_v|^3}) \frac{1}{\Omega-v_F p \cos(\theta-\theta_p)}
\end{equation}
Performing the $\theta$ integral (and take $v_F=1$ to simplify the expression), we find
\begin{equation}
\langle T_i T_i \rangle_v (\bm{p},\Omega)
\sim \int_{q_v}~ \int_{0}^{1} dl ~\frac{\Omega |q_v|}{p^2} \frac{\Omega}{p} \ln(1+\Omega \frac{|l|}{|q_v|^3}) \frac{1}{(1-(\Omega/p)^2)\sqrt{1+(q_v/v_f)^2-(\Omega/p)^2}}
\end{equation}
Since we are looking at the fermion near the Fermi surface, $q_v$ shall have a cut off which is smaller than $v_f$. Once we assume $q_v/v_F<1$,
\begin{equation}
\langle T_i T_i \rangle_v (\bm{p},\Omega)  \sim  \int^{\Lambda}_{0} d q_v~ \int_{0}^{1} dl ~\frac{\Omega |q_v|}{p^2} \frac{\Omega}{p} \ln(1+\Omega \frac{|l|}{|q_v|^3}) \frac{1}{(1-(\Omega/p)^2)\sqrt{1-(\Omega/p)^2}}
\end{equation}
where $\Lambda$ is the UV momentum cut off for $q_v$, integrate over $q_v$ and ignore the numerical constant,
\begin{align}
\langle T_i T_i \rangle_v (\bm{p},\Omega)&\sim  
  ~ \int_{0}^{1} dl ~\frac{\Omega}{p^2} \frac{\Omega}{p} (\Omega l)^{2/3} \frac{1}{(1-(\Omega/p)^2)\sqrt{1-(\Omega/p)^2}}\nonumber\\
&\sim  ~ \int_{0}^{1} dl ~\frac{\Omega^{8/3}}{p^3}( l)^{2/3} \nonumber\\
\end{align}
\end{widetext}
From the above calculation, it is obvious that the damping term from the vertex correction is of higher order compared to $ \frac{\Omega}{p}$.  One might worry that $\langle T_1 T_1 \rangle_v$ contains some singular damping term which cancels when sum with $\langle T_2 T_2 \rangle_v$. If this cancellation happens, then the damping coefficient $\langle T_1 T_1 \rangle_v$ must be a function of $\cos(4 \theta_p)$. We can then choose $ \theta_p=0$ and calculate $\langle T_1 T_1 \rangle_v$ to check the leading damping term. The leading order of damping for $\langle T_1 T_1 \rangle_v( \theta_p=0)$ is of order $\Omega^{5/3}/q^2$, which is still irrelevant compared to the damping term.

Thus, the vertex correction to the overdamped mode is irrelevant.

\providecommand{\noopsort}[1]{}\providecommand{\singleletter}[1]{#1}%
%

\end{document}